\documentclass[12pt]{iopart}

\newcommand{\bn}[1]{\mbox{\boldmath $#1$}} 
\usepackage{amssymb}
\usepackage{epsfig}
 
\begin{document}

\title[Energy spectrum and Landau levels in bilayer ...]
{Energy spectrum and Landau levels in  bilayer graphene 
with spin-orbit interaction}

\author{Francisco  Mireles$^{1,2}$ and John Schliemann$^{1}$}
\address{$^{1}$Institut f\"ur Theoretische Physik,
Universit\"at Regensburg, D-93049 Regensburg, Germany}
\address{$^{2}$Centro de Nanociencias y Nanotecnolog{\'{i}}a, 
Universidad Nacional Aut\'onoma de M\'exico, Ensenada, BC, C.P. 22800, M\'exico}
\ead{fmireles@cnyn.unam.mx}
\begin{abstract}
We present a theoretical study of the bandstructure and Landau 
levels in bilayer graphene 
at low energies in the presence of a transverse magnetic field and Rashba 
spin-orbit 
interaction in the regime of negligible trigonal distortion. Within an
effective low energy 
approach (L\"owdin partitioning theory) we derive an effective Hamiltonian 
for bilayer graphene  
that incorporates the influence of the Zeeman effect, the Rashba spin-orbit 
interaction, and 
inclusively, the role of the intrinsic spin-orbit interaction on the same 
footing. Particular 
attention is spent to the energy spectrum and Landau levels. Our modeling 
unveil the strong 
influence of the Rashba coupling  $\lambda_R $ in the spin-splitting of the 
electron and hole 
bands. Graphene bilayers with weak Rashba spin-orbit 
interaction show a 
spin-splitting linear in momentum and proportional to $\lambda_R $, but scales 
inversely 
proportional to the interlayer hopping energy $\gamma_1$. However, at robust 
spin-orbit 
coupling $\lambda_R $ the energy spectrum shows a strong warping behavior 
near the Dirac 
points. We find the bias-induced gap in bilayer graphene to be decreasing 
with increasing
Rashba coupling, a behavior resembling  a topological insulator transition. 
We further 
predict an unexpected assymetric spin-splitting and crossings of the Landau 
levels  due to 
the interplay between the Rashba interaction and the external bias voltage. 
Our results are 
of relevance for interpreting magnetotransport and infrared cyclotron 
resonance measurements, 
including also situations of comparatively weak spin-orbit coupling.
\end{abstract}

\maketitle

\section{Introduction}
Graphene and its bilayer (BLG) posses very distinctive physical 
properties.\cite{CastroNetoRMP09,Castro-Bilayer} Neglecting interactions, the 
low energy quasiparticles in pristine single and double layer graphene 
obey, respectively, linear and quadratic dispersion laws at the $K(K^{\prime})$ 
symmetry points.\cite{Abergela,Sarma,Peres} In the presence of a quantizing 
magnetic field $B>0$, the relativistic massless Dirac fermions of graphene 
exhibits  Landau levels (LLs) non-equidistant in energy, $E_n \propto \sqrt{|n| B }$, with $n=0,\pm1,\pm2,...$, the LL index.\cite{Haldane}The latter 
gives rise to the half integer quantum Hall effect at room 
temperature,\cite{Novoselov2} and to   the fractional quantum Hall effect at 
high magnetic fields in the presence of many body effects.\cite{Du, Bolotin} 
On the other hand, LLs for BLG show a rather intricate index sequence 
instead; with a roughly linear $B$-field dependence for low LLs, 
and a $\sqrt B$ for high LLs.\cite{Nakamura,Koshino,Vasilopoulus}At 
low energies,  the LLs in unbiased BLG follows the sequence 
$E_n \propto  \sqrt{|n|(|n|+1) }$ for $n\ge 1$ with a double degenerate 
zero-energy level, $E_o=0$ for $n=0$. \cite{McCann,Abergel-Falko,McCann-Review} 
Experimentally, the LLs dipole-allowed transition energies in single 
layer and BLG have been studied in detail by cyclotron 
resonance. \cite{Neugebauer,Henriksen-BLG,Henriksen-MLG} Most recently, 
phonon-mediated inter LL excitations have been explored  by 
magneto-Raman scattering experiments.\cite{Faugeras}

The particularly high interest in graphene spin physics
is strongly motivated by its expected prospects in nanoelectronics 
and spin-based devices for spintronics.\cite{Zutic-Fabian-DaSarma} In this 
realm, the role of the spin-orbit interaction (SOI) effects in graphene 
sheets is a phenomenon  under intense scrutiny. Two types of SOI in 
graphene have been identified; ({\it i}) the induced by  carbon 
intra-atomic SOI ({\it intrinsic}-SOI), and ({\it ii}), the SOI coming 
from the breaking of the space inversion symmetry of the hexagonal 
lattice ({\it extrinsic}-SOI), this is the so-called Rashba-SOI. 
The latter can have different origins, among those is the presence of a 
substrate, buckling, {\it ad}-atoms, or external electric fields. 

The magnitude of the excitation gap $\eta$ of the {\it intrinsic}-SOI 
in single layer is predicted to be very 
small (0.86-50\,$\mu$eV). \cite{Yao,McDonald-Group,HH,B-T,Gmitra,Sergei-Gmitra-Fabian,Konschuh}  
Likewise, estimates of the Rashba coupling  $\lambda_R$, leads to small 
values (few tens of $\mu$eV)  at typical electric fields 
($\sim$ 0.16\, mV/nm). \cite{HH,B-T,Gmitra} However recent spin-resolved
photoemission experiments in  graphene/Au/Ni(111)  showed enhancements of 
the Rashba coupling as large as 13 meV.\cite{Varykhalov} Induced distortions 
by neutral impurities ({\it ad-atoms})\cite{CN-G,Weeks} and  the interplay of 
buckling with Rashba-SOI also yield significant enhancement of the 
spin-splittings (up to 40\,meV).\cite{Samir} The role of the impurities and 
lattice deformations seems to be crucial for the observed long spin 
relaxation times (up to 2\,ns in BLG\cite{Barbaros}) linked to SOI 
effects\cite{Tombros, Jozsa}, as well as for  a non-monotonic dependence on bias voltage of the spin relaxation times in BLG due to the D'yakonov-Perel' spin-precession mechanism.\cite{Guido} 
Large spin-splittings ($\sim 0.22$ eV) of the graphene $\pi-$states attributed to Rashba-SOI has been reported also in epitaxial graphene on a Ni substrate.\cite{Dedkov} In theory, it has been shown that  the Rashba-SOI can induce significant changes in the bandstructure of graphene\cite{Rashba, Zaera} as well as in BLG.\cite{vanGelderen}

In graphene monolayers with Rashba coupling,  
the  Landau levels are described by 
$E{\scriptstyle^{(1)}_{n,\mu \pm}}=\mu\sqrt{  {\cal E}{\scriptstyle_{n,\pm  }^{(1)}} }$, 
(in units of $\lambda_R$), 
with\cite{Rashba} 

\begin{equation}   {\cal E}{\scriptstyle^{(1)}_{n, \pm}}=  
(2n-1)\tilde\Gamma^2+\frac{1}{2}
\left (1 \pm \sqrt{1+4(2n-1)\tilde\Gamma^2+4\tilde\Gamma^4} \right)
\label{Rashba-graphene-LL} 
\end{equation} 

\noindent for $n\ge 2$, where $\tilde\Gamma=\hbar v_F/l_B|\lambda_R|$,  
$v_F$ is the Fermi velocity ($\sim 10^6 $m\,s$^{-1}$), $l_B=\sqrt{c\hbar/eB}$ 
is the magnetic length, $\hbar$ the Plank's constant over $2\pi$, $c$ is the 
light velocity in vacuum, and $-e$ is the electron charge. Here $\mu=\pm$ 
stands for the  electron/hole LL branch. The lowest $n=0$ is given by $E_{0}=0$ 
whiles the $n=1$ level gives rise to three modes.  
A zero mode $E{\scriptstyle^{(0)}_{1}}=0$, in addition to its two nondegenerate 
modes at 
$E_{1\mu}= \mu\sqrt{1+2\tilde\Gamma^2}$. Exact solutions for the LLs in monolayer graphene under the influence of a Zeeman field and  spin-orbit interactions has been also reported recently by De Martino {\it et al.}.\cite{DeMartino} 

In this paper we show within low energy effective theory that for biased BLG in which the Rashba effect is the dominant SOI, its LLs  must follow  $E{\scriptstyle^{(2)}_{n,\mu \pm}}=\mu\sqrt{  {\cal E}{\scriptstyle_{n,\pm  }^{(2)}} }$, with 

\begin{equation}  
\label{LL-BLG}
{\cal E}_{n,\pm}^{(2)}=U^2+ \frac{n}{2} \left(\Gamma ^2 +2 n \, \omega ^2\pm \sqrt{4    
 \,\omega ^4+4 n \, \omega ^2 \Gamma ^2+\Gamma ^4}\right)
\end{equation}
 
\noindent for $n\ge 2$, with $U$  the interlayer bias energy, $\Gamma=2\sqrt{2}\lambda_R v_F\hbar/\gamma_1 l_B$, and $\omega=2v_F^2\hbar^2/\gamma_1 l_B^2$,  being $\gamma_1$   inter-layer hopping energy. As it occurs in graphene, in BLG with Rashba-SOI we obtain three modes for $n=1$, namely, the nondegenerate $  E ^{(2)}_{1\mu+}=\mu\sqrt{U^2+\Gamma^2 +2\omega^2}$ and $  E^{(2)}_{1-}=-U$, whereas for $n=0$ its eigenvalue also vanishes in the absence of gating, $E_{0-}^{(2)}=U$.   Eq.(\ref{LL-BLG}) comprises one of the main results of this contribution.

The aim of this study is to investigate the energy spectrum and the Landau levels in bilayer graphene under the influence of sizable spin-orbit interactions (SOI) of both, intrinsic and Rashba types. An effective low energy Hamiltonian for bilayer graphene  that includes  both types of SOI and Zeeman effect is derived within L\"owdin partitioning theory. Whiles the Rashba SOI in single layer graphene is known to modify its otherwise conic spectrum, to a spectrum that includes two zero gap bands and two gapped branches of width $2\lambda_R$ (with parabolic shape, similar to unbiased BLG);\cite{Rashba} here in contrast, the (unbiased) BLG with Rashba-SOI shows a spin-splitting which is linear in momentum and proportional to $\lambda_R $, but inversely proportional to the interlayer hopping parameter $\gamma_1$. We predict also a strong influence of the Rashba spin-orbit interaction in the warping of the low energy bandstructure of biased bilayer at comparatively weak spin-orbit coupling $\lambda_R $. It is found that the bias-induced gap in bilayer graphene decreases as the Rashba strength coupling is increased. It is further predicted the rise of an unexpected asymmetric spin-splitting of the Landau levels due to the interplay among the Rashba coupling and the external bias voltage.  

The remainder of the paper is organized as follows. In Sec.\,{\bf 2} we outline the model and derivation for the low energy BLG effective Hamiltonian. The Landau level spectrum in the presence of Rashba-SOI is discussed in Sec.\,{\bf 3}. The band spectrum properties and the Landau levels of BLG with Rashba-SOI are analyzed in detail in Sec.\,{\bf 4} and Sec.\,{\bf 5}. A summary of our results is given in Sec.\,{\bf 6} . Additionally, we provide three appendixes. In Appendix A we outline the derivation (L\"owdin partitioning theory) of the low energy Hamiltonian in the presence of intrinsic and Rashba types of SOI, as well as the Zeeman effect. In Appendix B a detailed description of the eigenvalues of the low energy Hamiltonian is given, and finally in Appendix C the Landau levels for BLG with Rashba-SOI are derived.   

\section{Low energy effective Hamiltonian for bilayer graphene}

Here we focus in the derivation of the low energy effective Hamiltonian for BLG with SOI in the presence of magnetic field that eventually leads to Eq.(\ref{LL-BLG}). We start by considering a pile of two graphene layers (BLG) in which the sites $A_2$ of the upper layer 2 lies directly on top of the $B_1$ sites of the bottom layer 1 (AB-Bernal stacking). At the vicinity of the Dirac $K$ symmetry point, the effective non-interacting bilayer graphene Hamiltonian $H_o$, written in terms of the spin-dependent basis  
$|\Psi^{\dagger}_{K}\rangle =\{\psi_{A_{1\uparrow}}, \psi_{A_{1\downarrow}},\psi_{B_{1\uparrow}}, \psi_{B_{1\downarrow}},\psi_{A_{2\uparrow}}, \psi_{A_{2\downarrow}},\psi_{B_{2\uparrow}}, \psi_{B_{2\downarrow}} \}$,  has the $8\times8$ matrix form

\begin{equation}
 H_o=\left(
\begin{array}{cc}
 H_+ & V_1 \\
 V_1^{\dagger} & H_-
\end{array}
\right);\,\,\,H_{\pm}= \left(
\begin{array}{cccc}
 \pm U& 0 & \gamma\,  \pi^{\dagger} & 0 \\
 0 & \pm U   & 0  & \gamma\,  \pi^{\dagger} \\
 \gamma \, \pi & 0  & \pm U   & 0 \\
 0 & \gamma \, \pi  & 0 & \pm U  
\end{array}
\right),
\end{equation}

\noindent where  $\pi=\pi_x+i\pi_y$,  with  ${\bn \pi}=\bn p+e \bn A/c=(\pi_x,\pi_y)$ is the canonical momentum, and $\bn A$ is the vector potential. 
Here   $\gamma \equiv v_F=\gamma_o a\sqrt{3}/2\hbar$, with $\gamma_o\sim 3.1$ eV (intra-layer hopping energy) and $a=0.246$ nm is the lattice parameter.\cite{CastroNetoRMP09} The electrostatic potential $\pm U$ of the bottom/upper layer is gate voltage adjustable and opens a gap of $2U$ in the spectrum.\cite{Castro-Bilayer} The dominant interlayer interaction in BLG is described to first approximation by the term

\begin{equation} 
\label{V1}
V_1=\left(
\begin{array}{cccc}
 -v_4\pi^{\dagger}  & 0 & \gamma _1 & 0 \\
 0 & -v_4\pi^{\dagger} & 0 & \gamma _1 \\
 v_3\pi & 0 & -v_4\pi^{\dagger} & 0 \\
 0 & v_3\pi  & 0 & -v_4\pi^{\dagger}
\end{array}
\right),
\end{equation}

\noindent where $\gamma_1$  is the nearest neighbor (interlayer) hopping energy ($\sim 0.1\gamma_o$). The terms proportional to the velocities $v_3=\gamma_3 a\sqrt{3}/2\hbar$ and $v_4=\gamma_4 a\sqrt{3}/2\hbar$  arise due to second nearest neighbor (interlayer) hopping processes associated with $\gamma_3$ and $\gamma_4$ tight-binding parameters, respectively.\cite{McCann-Review,Kusmenko} The former produces a trigonal warping whose characteristic energy is $E_3=\frac{1}{2}m^*v_3^2=\gamma_1(\gamma_3/2\gamma_o)^2\sim$ 1 meV, being $m^*=\gamma_1/2v_F^2$ the electron effective mass \cite{McCann-Review}, whiles the latter has yet a smaller characteristic energy, $E_4=\frac{1}{2}m^*v_4^2=\gamma_1(\gamma_4/2\gamma_o)^2\sim$ 0.2 meV. Since typically $E_{3,4}\ll \gamma_1$ in BLG, the contributions in Eq.(\ref{V1}) coming from the terms with  velocities $v_3$ and $v_4$ are negligible. 
However we should notice that, when considering Rashba spin-orbit interaction, there might be situations in which $v_3$ and $v_4$ may be important, particularly at very weak Rashba-strengths of the order of $E_{3,4}$.  We would like to mention nevertheless, that there is experimental (and theoretical) evidence of  Rashba-SOI spin-splittings of one order of magnitude \cite{Varykhalov,CN-G,Samir} and even larger \cite{Dedkov} than the typical distortion energies $E_{3,4}$. In any case, its inclusion in the model can be readily  incorporated in the general derivation of the reduced effective Hamiltonian via the L\"owdin perturbation theory described in Appendix A. The regime in which there is a possible interplay among the $v_{3,4}$ terms and  Rashba SOI is out of the scope of the present analytical study and for sake of clarity and simplicity these trigonal terms will be disregarded hereafter. 

When taking into account explicitly the presence  of  {\it intrinsic-SOI}, Rashba-SOI and   Zeeman splitting, the total eight-band effective Hamiltonian will read  

\begin{equation}
\label{TotH}
H_K=H_o + H_{R}+H_{I}+H_Z.
\end{equation}

\noindent The second term to the right in Eq.(\ref{TotH}) arises due the influence of an  effective electric field  perpendicular to the BLG plane  producing a Rashba type of SOI. The the leading contribution to the Rashba-SOI Hamiltonian, $H_R$, is modeled  as follows:  

\begin{eqnarray} 
 H_R= \left( 
\begin{array}{cccc} 
 0  &   i \lambda_R \,\sigma_- &  0 &  0 \\ 
 -i \lambda_R\, \sigma_+ & 0 & 0  &  0 \\
 0& 0  & 0 & i \lambda_R\, \sigma_- \\
0 & 0 & -i \lambda_R\, \sigma_+ & 0
\end{array}
 \right)  ;  
\end{eqnarray}

\noindent here $\lambda_R $ parameterize the strength of the intra-layer Rashba-SOI, as in monolayer graphene, with $\sigma _{\pm}=\frac{1}{2}\left(\sigma _x\pm i\sigma_y\right)$, being $(\sigma_x,\sigma_y)$ the usual $2\times 2$ Pauli spin matrices. The intensity of the Rashba-SOI can be sizable ($\lambda_R \sim 10$ meV) due for instance to the presence of a metallic substrate.\cite{Varykhalov} 
Within tight-binding theory, it is understood that the Rashba-SOI arises because of the effective nearest-neighbor hopping of two $p_z$ orbitals with opposite spins under the presence of an applied transverse electric field.\cite{Sergei-Gmitra-Fabian} 

Recently it has been predicted that $\lambda_R $ can be even larger (of a few tens of meV) due to buckling effects in conjunction with external electric fields.\cite{Samir} Furthermore by varying the electric field the Rashba parameter can be tuned.   A possible inter-layer Rashba spin-orbit coupling of strength $\lambda_R^{\perp}$ can  in principle  be present in BLG as well, however such contributions will be ignored here because of its predicted slight influence on the energy bands for $\lambda_R^{\perp}/\gamma_o \lesssim 0.3$. \cite{vanGelderen}

Additionally, in the same basis set above, the {\it intrinsic}-SOI Hamiltonian for BLG   (third term to the r.h.s. in Eq.(\ref{TotH}) ) shall follows the $8\times 8$ matrix form 

\begin{equation} 
H_{I }=\left(
\begin{array}{cccc}
 \eta \,s_z & 0 & 0 & 0 \\
 0 & -\eta\, s_z & 0  &  0 \\
 0& 0  & \eta \,s_z & 0  \\
 0 & 0 & 0 & -\eta\, s_z
\end{array}
\right),
\end{equation}

\noindent  with $\eta$ the intrinsic SOI constant and $s_z$ is the spin operator along the $z$-axis, perpendicular to the BLG plane. 
The intrinsic SOI is a second order tunneling process (within tight-binding theory), which involves next-nearest-neighbor hopping events of $p_z$ electrons of a given spin. 
As mentioned in the introduction, the value of the excitation energy $\eta$  has been inferred to be very small in monolayer graphene  in both $K(K')$ symmetry points ($\lesssim 50$\,$ \mu$eV), even if one consider interactions up to first order by including the unoccupied $d$ and higher orbitals.\cite{Gmitra,Sergei-Gmitra-Fabian} Interestingly, in BLG, taking into account the interlayer overlapping of the $\pi$ and $\sigma$ bonds yields  enhanced values of the {intrinsic}-SOI; about one order of magnitude larger than in single layer graphene ( $\sim0.1$ meV).   \cite{Guinea} We would like to emphasize here that  such values   still somewhat weak, compared with those relatively large strengths, of which reportedly, the Rashba parameter $\lambda_R$ can acquire (similar to the values attained in III-V semiconductors). Nevertheless, for  generality, the intrinsic SOI has been incorporated  in the present derivation of the low energy effective Hamiltonian. This will be helpful when considering BLG in the extreme limit, {\it i.e.} when the intrinsic-SOI $\eta$ is much stronger than the Rashba-SOI  $\lambda_R$ parameter ($\eta\gg \lambda_R$). However, we shall concentrate here our discussion on the bandstructure and Landau levels of BLG for the case $\lambda_R\gg \eta$, the opposite limit will be treated elsewhere.  

The last term in Eq.(5) arises if an external magnetic field $B$ is present, affecting the energetic of the quasiparticles in the form of a Zeeman interaction, $H_Z$, for a field perpendicular to the BLG plane it reads

\begin{equation}
 H_Z=\Delta\,( {\bn I}\otimes\sigma_z),
\end{equation}

\noindent where ${\bn I}$ is a $4\times4$ unit matrix, $2\Delta=  g \mu_B B$ is the Zeeman splitting energy,   $g$ is the electron Land\'e factor, and $\mu_B$ is the Bohr magneton. Note that even at relatively high magnetic fields ($B=10$ T), the Zeeman splitting it is still somewhat small ($\Delta\sim 1.1$\,meV), whiles it is practically negligible at low fields. Note that the condition, 

\begin{equation}
\eta\lesssim\Delta\ll \lambda_R\lll \gamma_1 ,
\end{equation}

\noindent typically holds at finite fields ($B\gtrsim 0.1$ T). This condition will allow us to work safely within the low energy theory and derive an effective Hamiltonian for BLG, including  the extrinsic (Rashba)-SOI, intrinsic-SOI, and Zeeman effect on the same footing. We finally should remark that the total Hamiltonian (\ref{TotH}) is valid near $K$ symmetry point only. For the $K^{\prime}$ point of the Brillouin zone, the Hamiltonian  $H_{K^{\prime}}=\Sigma_y H_K \Sigma_y^{-1}$, with $\Sigma_y=\sigma_y\otimes {\bn I}$, should be used instead.

\subsection{Low energy bilayer Hamiltonian}
Using L\"owdin partitioning theory \cite{van-Vleck,Cohen} the full $8\times 8$ Hamiltonian $H_K$ can be projected through a canonical transformation\cite{Zhang} into a $4\times4$ low energy effective Hamiltonian ${\cal H}$  in an appropriate basis (see Appendix A). The projected low energy Hamiltonian  can be further expressed in terms of Kronecker products of $2\times2$ matrices   and conveniently separated into the sum of the Hamiltonians  (keeping terms up to $1/\gamma_1^2$),  

\begin{equation}\label{TotalH}
{\cal H}={\cal H}^{(o)}+{\cal H}^{(1)}+{\cal H}^{(2)}+\mathcal{O}(1/\gamma^3),
\end{equation}
\noindent in which the  term independent of the interlayer hopping parameter $\gamma_1$  reads, 
\begin{equation}
{\cal H}^{(o)}= -\sigma_z\otimes(U\sigma_o+\Delta \sigma_z) - \eta\, (\sigma_o\otimes\sigma_z), 
\end{equation}

\noindent where   $\sigma_z$ is the $z-$component of the Pauli matrices and  $\sigma_o$ is the $2\times2$ unit matrix. The dominant  contribution to ${\cal H}$ is described by the Hamiltonian ${\cal H}^{(1)}$, given by

\begin{equation} 
\label{H1}
{\cal H}^{(1)}    =    -\frac{\gamma ^2}{\gamma _1}\left(\! 
\begin{array}{cc}
 0 & \left(\pi^\dagger\right)^2 \nonumber\\ 
 \pi ^2 & 0
\end{array}
\! \right)\otimes \sigma_x   
 +  \frac{2i\lambda_R\gamma }{\gamma _1}\left(\! 
\begin{array}{cc}
 0 & -\pi^\dagger \\ 
 \pi  & 0
\end{array}
\! \right)\otimes   s_+,  
\end{equation}

\noindent where we have defined the operator $s_{\pm}=\frac{1}{2}(\sigma_o \pm \sigma_z)$. 
Without Rashba-SOI ($\lambda_R=0$), Eq.(\ref{H1}) decouples to the usual effective BLG Hamiltonian obtained within low energy theory in the absence of trigonal warping effects.\cite{McCann} Such term gives rise to well known parabolic spectrum of the massive Dirac quasiparticles in BLG. The second term in ${\cal H}^{(1)}$ is linear in momentum and can be viewed as a renormalization of the Rashba coefficient due to the presence of the higher bands. Notice that it scales as the inverse of the interlayer hopping energy $\gamma_1$.

The remaining terms proportional to $1/\gamma_1^{2}$ in Eq.(\ref{TotalH}) are compacted into the sum  ${\cal H}^{(2)}=\sum_{i=1}^{4} h^{(2)}_i$, with

\begin{eqnarray}  
h^{(2)}_1   &=&  \frac{2U \gamma ^2}{\gamma _1^2}\left(\!\!
\begin{array}{cc}
 \pi^\dagger\pi  & 0   \\ 
 0 & -\pi \pi^\dagger
\end{array}
\!\!\right)\otimes \sigma_o \,,  \nonumber \\ 
h^{(2)}_2    &=&   \frac{U \lambda _R{}^2}{\gamma _1^2}(\sigma_z\otimes  s_+)+  \frac{(\Delta+\eta)\lambda_R^2}{\gamma _1^2}(\sigma_o\otimes s_+) \,, \nonumber \\
h^{(2)}_3    &=&  \frac{i(2U + \Delta)\lambda _R }{\gamma _1^2}  \left(\!\! 
\begin{array}{cc}
 \pi  & 0 \\
 0 & \pi^\dagger
\end{array}
 \!\!\right) \otimes  \sigma_+ + h.c.  \,,  \nonumber\\
 h^{(2)}_4   &=&   -\frac{i\Delta\lambda_R }{\gamma _1^2}s_{-}\otimes \left(\!\! 
\begin{array}{cc}
 0 & -\pi^\dagger \\
 \pi  & 0
\end{array}
\!\! \right)+\frac{i\eta\lambda_R }{\gamma _1^2}s_+\otimes \left(\!\! 
\begin{array}{cc}
 0 &  \pi  \\
 -\pi^\dagger  & 0
\end{array}
\!\! \right).  \nonumber 
\end{eqnarray}

The low energy effective Hamiltonian ${\cal H}$ described in Eq. (\ref{TotalH}) is valid within the energy range  $\epsilon \lesssim \gamma_1 $. Notice that  it will be fairly sufficient to keep only the leading order contribution $h^{(2)}_1$ in ${\cal H}^{(2)}$ given the typical smallness  of the ratios $\lambda_R^2/\gamma_1^2$, $\lambda_R\Delta/\gamma_1^2$, and $\lambda_R\,\eta/\gamma_1^2$ appearing in ${\cal H}^{(2)}$  together with the assumption  $U<\gamma_1$. Hence the description for the low energy (and momentum) effective Hamiltonian will be given by ${\cal H}={\cal H}^{(o)}+{\cal H}^{(1)} + h^{(2)}_1$.
  
If we further neglect the Zeeman and the intrinsic SOI ($\Delta=\eta =0$) the effective bilayer Hamiltonian with Rashba-SOI written in the atomic basis  $\{\psi_{A_{2\uparrow}}, \psi_{A_{2\downarrow}},\psi_{B_{1\uparrow}}, \psi_{B_{1\downarrow}} \}$ reduces to

\begin{equation} 
\label{Heff}
{\cal H}=\left(
\begin{array}{cccc}
-{U}+ \xi\pi^{\scriptscriptstyle\dagger} \pi & 0 & -\beta   \left(\pi^{\scriptscriptstyle\dagger}\right)^2  & 0\\
 0 & -{U} +\xi\pi^{\scriptscriptstyle\dagger} \pi& -i \alpha\, \pi^{\dagger}   &  -\beta   \left(\pi^{\dagger}\right)^2 \\
 -\beta \, \pi^2  & i \alpha \, \pi    & {U} -\xi\pi \pi^{\scriptscriptstyle\dagger} & 0 \\
 0 & -\beta \, \pi^2 & 0 & {U} -\xi\pi \pi^{\scriptscriptstyle\dagger}
\end{array}
\right),
\end{equation}

\noindent where we have defined the parameters  $\xi=2U\tilde \gamma^2$, $\alpha = {2\tilde\gamma\,\lambda _R }$ and $\beta = \gamma_1{\tilde\gamma ^2}$, with $\tilde\gamma=\gamma/\gamma_1$. 

\subsection{Bilayer graphene spectrum with Rashba effect at zero field }

Without magnetic field ($B=0$), $\pi^{\dagger} = \hbar k_{-}=\hbar(k_x - ik_y)$ and $\pi = \hbar k_{+}=\hbar(k_x + i k_y)$ and the eigenvalues of Eq.(\ref{Heff}) are readily determined  by (Appendix B)

\begin{equation}
\label{Eko}
\varepsilon _{\mu s}(k)=\frac{\mu}{2} \sqrt{4\, (U-\xi k^2)^2+k^2\left(\sqrt{\alpha^2+4 k^2 \beta^2}-s\, \alpha \right)^2 },
\end{equation}
 
\noindent with $k=\sqrt{k_x^2+k_y^2}$. Here $\mu=\pm$ describe the electron/hole branch, whiles $s=\pm$ characterizes its spin chirality.  
Therefore, the low-energy spectrum consist of four spin-splitted bands, two conduction and two valence bands. For the unbias voltage case ($U=0$) the spectrum reduces simply to
\begin{equation}
\label{EUnbias}
\varepsilon^o _{\mu s}(k)=\frac{\mu}{2} k\left(\sqrt{\alpha^2+4 \beta^2 k^2 }-s\, \alpha \right). 
\end{equation}

\begin{figure}[h]
\vspace{-0.35in}
\hspace*{4.0cm}\includegraphics[width=4.0in,height=4.0 in]{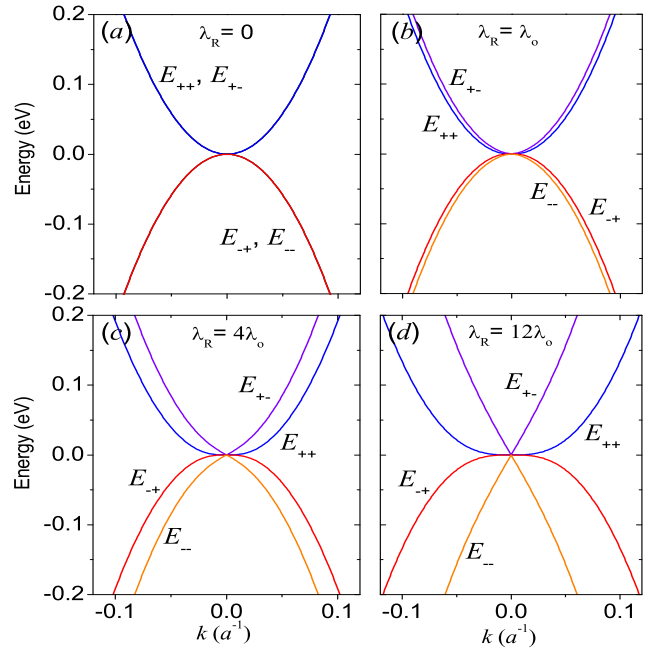}
\vspace{-0.4in}
\caption{ (Color online) Low quasiparticle energy spectrum for bilayer graphene with Rashba-SOI. The spin degeneracy of the bands at $\lambda_R=0$ (a) is lifted for $\lambda_R\ne0$ (b)-(d). As the strength of the Rashba parameter is increased, the symmetry of the bands is broken producing a cone chape for the innermost bands (s=-) at high values of $\lambda_R$.    }
\label{Dispersion-unbias}
\end{figure}

We note that in contrast with single layer graphene,\cite{comment1} in BLG the Rashba-SOI it is induced a linear spin-splitting in momentum of the conduction and valence bands, in close analogy with the Rashba interaction arisen in two-dimensional electron gases in semiconductor heterostructures. In addition we observe that the cyclotron effective mass at the Fermi energy $m_c^{(s)}=k/(\partial\,\varepsilon^o _{\mu s}/\partial \,k )$ turns out to be spin-dependent as long as $\lambda_R\ne 0$ following the relation, 

\begin{equation}
 m_c^{(s)}=m^*\frac{4\mu \sqrt{\pi n } \lambda_{\beta}\sqrt{1+4\pi n\lambda_{\beta}^{2}}}{1+8\pi n \lambda_{\beta}^2-s\sqrt{1+4\pi n\lambda_{\beta}^{2}} },
\end{equation}

\noindent with $\lambda_{\beta}=\beta/\alpha=\gamma/2\lambda_R$, and we have expressed the Fermi wave number in terms of the 2D carrier density via $k_F=\sqrt{\pi n}$. Notice that in the limit case $\lambda_{\beta}\rightarrow \infty$, the cyclotron mass $m_c^{(s)}\rightarrow \mu m^*=\mu \gamma_1/2v_F^2$, as one expects  for unbiased BLG in the absence of SOI. However as the carrier density $n\rightarrow 0$ the cyclotron effective mass does not diverges as predicted by tight-binding models.

The normalized eigenvectors $|\psi _{ks }^{(\mu)}\rangle$ of ${\cal H}$ corresponding to the electron and hole bands $\mu=\pm$, respectively, for the case of spin up ($s=+$) are written in the four-vector form (Appendix B) as,
  
\begin{equation}
\label{eigen1} 
|\psi _{k+ }^{(\mu)}\rangle=\frac{1}{\sqrt{1+{(\chi}_{ \scriptscriptstyle +}^{\scriptscriptstyle{(\mu)}})^2}}\left(
\begin{array}{c}
 -i\, e^{-2i \phi }\sin (\theta /2) {\chi}_{ \scriptstyle +}^{\scriptstyle{(\mu)}}\\
 e^{-i \phi }\cos (\theta /2){\chi}_{ \scriptstyle +}^{\scriptstyle{(\mu)}} \\
 -i\,\cos (\theta /2) \\
 i\,e^{-i \phi }\sin (\theta /2)
\end{array}
\right) ,
\end{equation}

\noindent whiles the normalized eigenvectors for the spin down ($s=-$) electron/hole bands are specified by

\begin{equation} 
\label{eigen2}
|\psi _{k- }^{(\mu) } \rangle=\frac{1}{\sqrt{1+{(\chi}_{ \scriptscriptstyle -}^{\scriptscriptstyle{(\mu)}})^2}}\left(
\begin{array}{c}
 i\, e^{-2i \phi }\cos (\theta /2) {\chi}_{ \scriptstyle -}^{\scriptstyle{(\mu)}}\\
 e^{-i \phi }\sin (\theta /2){\chi}_{ \scriptstyle -}^{\scriptstyle{(\mu)}} \\
 i\,\sin (\theta /2) \\
 i\,e^{-i \phi }\cos (\theta /2)
\end{array}
\right) , 
\end{equation} 

\noindent in which we have defined the dimensionless parameter

\begin{equation}
\label{R}
{  \chi}_{ \scriptstyle s}^{\scriptstyle{(\mu)}}=\frac{ { U-\xi k^2} +\mu \sqrt{ {\cal R}_{\mu s}^2 + {(U-\xi k^2)}^2} } { {\cal R}_{\mu s}  },  
\end{equation}

\noindent with  $  \theta  = \tan^{-1}(2\beta  k/\alpha)$, and  $\phi$   is the azimuthal angle  of the in-plane wave vector, ${\bn k}=k(\cos\phi,\sin\phi)$. The denominator of Eq. (\ref{R}) is explictly, ${\cal R}_{\mu s}=\mu\,\varepsilon^o _{\mu s}(k)=|\varepsilon^o _{\mu s}(k)|$, which  implies  ${\cal R}_{+ s}={\cal R}_{- s}$, and therefore the relation ${  \chi}_{ \scriptscriptstyle \sigma}^{\scriptscriptstyle{(+)}} {  \chi}_{ \scriptscriptstyle \sigma}^{\scriptscriptstyle{(-)}}=-1$ it is always satisfied. Without external bias voltage (${U}=0$),   the parameter ${  \chi}_{ \scriptstyle \sigma}^{\scriptstyle{(\pm)}}$ reduces to $\pm 1$ for all $k$. 

\subsubsection{Spin and valley polarization}
The expectation value of the valley (charge) polarization and spin orientation  are defined   as $\langle {\bn \tau}   \rangle_{\mu s}  =\langle \psi_{ks }^{(\mu)}| \bn \tau | \psi _{ks }^{(\mu)}  \rangle$, and  $\langle {\bn S}   \rangle_{\mu s}  =\langle \psi_{ks }^{(\mu)}| \bn S | \psi _{ks }^{(\mu)}  \rangle$, respectively. Here the valley and spin operators are ${\bn \tau = {\bn \sigma}\otimes\sigma_o}$ and $\bn S= \frac{\hbar}{2}(\sigma_o\otimes {\bn \sigma})$, with ${\bn \sigma}=(\sigma_x,\sigma_y,\sigma_z)$ the vector of Pauli matrices. Using the results of Eqs. (\ref{eigen1}) and (\ref{eigen2}) the components for the charge polarization leads to the expressions

\begin{eqnarray}
\langle { \tau_x}   \rangle_{\mu s}  & = & \frac{2{\chi}_{ \scriptstyle s}^{\scriptstyle{(\mu)}}}{({\chi}_{ \scriptstyle s}^{\scriptstyle{(\mu)}})^2+1} \sin \theta \cos (2 \phi), \\
\langle { \tau_y}   \rangle_{\mu s}  & = &\frac{2{\chi}_{ \scriptstyle s}^{\scriptstyle{(\mu)}}}{({\chi}_{ \scriptstyle s}^{\scriptstyle{(\mu)}})^2+1} \sin \theta \sin (2 \phi), \\
\langle { \tau_z}   \rangle_{\mu s}  & = & \frac{({\chi}_{ \scriptstyle s}^{\scriptstyle{(\mu)}})^2-1}{({\chi}_{ \scriptstyle s}^{\scriptstyle{(\mu)}})^2+1}, \\ \nonumber
\end{eqnarray}

\noindent whereas  the components of the spin polarization (in units of $\hbar/2)$ satisfy  
\begin{eqnarray}
\langle { S_x}   \rangle_{\mu s}  & = &   -s\sin \theta \sin ( \phi), \\
\langle { S_y}   \rangle_{\mu s}  & = &  + s\sin \theta \cos (\phi), \\
\langle { S_z}   \rangle_{\mu s}  & = &   s\frac{1-({\chi}_{ \scriptstyle s}^{\scriptstyle{(\mu)}})^2}{1+({\chi}_{ \scriptstyle s}^{\scriptstyle{(\mu)}})^2}\cos \theta.  
\end{eqnarray}

As it occurs with the standard Rashba SOI in semiconductors, in BLG the orientation of the spin-polarization in the plane  is always perpendicular to the direction of the momentum,   $\langle {{\bn S} }   \rangle \cdot {\bn k}=0$. We notice also that, in contrast with single layer graphene, in BLG the  dot product $\langle {{\bn S} }   \rangle \cdot {\bn \tau}\ne 0$ in general. Interestingly,  as long there is a bias voltage present ($ U\ne 0$), both the spin and valley polarization have a nonzero component out of the BLG plane (along the $z-$axis). However,  the absence of bias voltage (${ U}= 0$) the amplitude of the charge and spin polarization develops $k-$dependent oscillations. Explicitly, $|\langle {\bn \tau}\rangle |=|\langle {{\bn S} }\rangle|=\sin\theta$ for all $|{\bn k}|\ne0$, and vanishes at ${\bn k}=0$; in close analogy with the known result in single layer graphene.\cite{Tobias-Schliemman} From Eqs.(22)-(24) the spin-polarization in the unbias configuration can compactly written as

\begin{equation}
\label{S}
\langle {{\bn S} }\rangle_{\mu s}=   \frac{2 s \beta }{\sqrt{\alpha^2+4\beta^2k^2}}(\hat z  \times {\bn k}), 
\end{equation}  

\noindent that is, $\langle {{\bn S} }\rangle_{\mu s}$ is forced to lay on the BLG plane. Clearly, as the Rashba SOI coefficient $\lambda_R \rightarrow  0$, {\it i.e.} $\alpha \rightarrow  0$, the magnitude of the spin-polarization reaches it maximum value, $|\langle{\bn S} \rangle_{\mu s}| \rightarrow   1$ as the electron/hole spin is conserved. Finally, we notice that Eq. (\ref{S}) for the spin orientation  in unbiased BLG is formally identical to that obtained in single layer graphene with Rashba SOI.\cite{Rashba} 

\section{ Landau levels in bilayer graphene with Rashba SOI}

For a magnetic field $B\ne 0$ perpendicular to the BLG plane, the operators $\pi$ and $\pi^{\dagger}$ do not commute any more since its components fails to do so, and of course, care has to be exercised in their ordering. By making the substitution in Eq. (\ref{Heff}) of the momentum operators in terms  of the  Bose operators, $\pi^{\dagger}=\sqrt{2}\hbar\,a^{\dagger}/l_B$ and $\pi=\sqrt{2}\hbar\,a/l_B$  with $\left[a,a^{\dagger}\right]=1$, the effective Hamiltonian in the limit of low bias ($U\ll\gamma_1$) takes the form

\begin{equation} 
{\hat H}_L =-\left(
\begin{array}{cccc}
 {  U} & 0 & \omega \left(a\right)^2  & 0 \\
 0 & {  U} & i\, \Gamma  a   & \omega \left(a\right)^2 \\
 \omega\,  (a^{\dagger})^2  & -i\, \Gamma  a^{\dagger} & -{  U} & 0 \\
 0 & \omega\,  (a^{\dagger})^2 & 0 & -{ U}
\end{array}
\right)
\end{equation}
 
\noindent with the notation $\Gamma=\sqrt{2}\hbar\,\alpha/l_B$ and $\omega=2  \hbar^2\beta/l_B^2$. The eigenfunctions of ${\hat H}_L$ can now be written in the form, $| \psi_n \rangle=(c_1^{(n-2)}|n-2\rangle,c_2^{(n-1)}|n-1\rangle,c_3^{(n)}|n\rangle,c_4^{(n+1)}|n+1\rangle)^T$, where  $|n\rangle \equiv \xi_n$ are the usual harmonic oscillator eigenfunctions satisfying $a^{\dagger}|n\rangle=\sqrt{n+1}|n+1\rangle$ and  $a|n\rangle=\sqrt{n}|n-1\rangle$. Consequently one can write the expectation value  
$\langle\psi_n|{\hat H}_L|\psi_n\rangle =\langle\phi_n|{\cal H}_n |\phi_n\rangle$, with   

\begin{equation} 
\label{HLandau}
{\cal H}_n=\left(
\begin{array}{cc}
  - { U} \sigma_o & {\cal F}_n \\
 {\cal F}^{\dagger}_n & {  U} \sigma_o
\end{array}
\right),
\end{equation}
\noindent and $|\phi_n\rangle =  (c_1^{(n-2)},c_2^{(n-1)},c_3^{(n)},c_4^{(n+1)} )^T$ satisfying the normalization condition $\langle\phi_n|\phi_n\rangle=1$, whiles
\begin{equation} 
{\cal F}_n=-\left(
\begin{array}{cc}
    \omega \sqrt{n(n-1)}& 0 \\
   i\, \Gamma \sqrt{n}  &  \omega \sqrt{n(n+1)}
\end{array}
\right).
\end{equation}
 
 The eigenvalues of (\ref{HLandau})  leads to the Landau spectrum given by Eq. (2), which, in the absence of a bias gate voltage across the layers ($U=0$), reads (Appendix C)

\begin{equation} 
\label{LL0}
\varepsilon _{n,\mu\pm }^o =\frac{\mu}{\sqrt{2}}  \sqrt{n \Gamma^2 + 2 n^2 \omega^2 \pm n \sqrt{4 \omega ^4+4 n \omega ^2 \Gamma ^2+\Gamma ^4}  }, 
\end{equation} 
\noindent for $n\ge2$. In order to gain further physical insight of the behavior of the Landau levels in BLG, it is illustrative to analyze the limit cases at zero, weak(large) Rashba-SOI relative to the magnetic field strength, with and without bias voltage.

\subsection{Approximate solutions for $U=0$.}

({\it i}) {\it Zero Rashba-SOI}  ($\Gamma=0$). In the vanishing Rashba-SOI strength limit, Eq. (\ref{LL0}) reduces to $\varepsilon _{n,\mu \pm }^o
=\mu\sqrt{n(n\pm1)}\,\omega_oB$,  with $\omega_o = e\hbar/m^*c$, and coincides with the LL spectrum (double degenerate in spin) reported in the literature for BLG in the absence of Rashba-SOI. The linear response with {\it B} stems from the parabolic dispersion laws of BLG for this case. \\

\noindent ({\it ii}) {\it Weak Rashba-SOI}  ($\Gamma/\omega\ll 1$). At very weak Rashba-SOI strength values (large fields), the LL's still evolve approximately linear with {\it B}, but shifted by a small energy proportional to $\lambda_R^2$, described by 
\begin{equation}
\varepsilon _{n,\mu \pm }^o\simeq\mu \left( \omega_o B \pm \frac{\Gamma_o^2}{4\omega_o}\right)\sqrt{n(n\pm1)},\quad \mu=\pm 
\end{equation}

\noindent with $\Gamma_o =  \sqrt{2 \omega_o/m^*}( \lambda_R/v_F )$ and $n\ge2$. Since $\Gamma_o^2/4\omega_o=\lambda_R^2/\gamma_1$,  only for rather large Rashba SOI coefficient ($\lambda_R\simeq\gamma_1$) strengths gives rise to significant broken degeneracies of the electron/hole  LLs as described next in the opposite regime. \\

\noindent ({\it iii}) {\it Strong Rashba-SOI}  ($\Gamma/\omega \gg 1$) Alternatively, in the very strong Rashba-SOI limit (small fields), the LL level spectrum is well described by  
$\varepsilon _{n,\mu + }^o \simeq \mu  \,  \sqrt{ n }\,\Gamma=\mu  \, \Gamma_o\sqrt{ n B}$, and $\varepsilon _{n,\mu - }^o \simeq \mu(\omega_o^2/\Gamma_o)\sqrt{n(n^2-1)}B^{3/2} $, with $n\ge2$. The change in the field dependence from $B^{1/2}$ of the $\nu=+$ chiral states to $B^{3/2}$ for the $\nu=-$ states is unique in BLG and give rise to a multiplicity of LL crossings as shown later.

In addition, the energy spectrum with LL level index $n=1$ and $n=0$ are special cases, giving rise to three levels, one at zero energy ($\epsilon_{0-} =0$) and two nondegenerate levels for $n=1$. In the high field limit, $\Gamma/\lambda_R\gg 1$, we get $E_{1\mu+}\simeq\mu\sqrt{2}(\Gamma_o^2/4\omega_o + \omega_o B)$, whiles $E_{1\mu-}\simeq\mu\Gamma_o\sqrt{B}$ in the weak field regime.

\subsection{Approximate solutions for $U \ne 0$.}

({\it i}) {\it Zero Rashba-SOI in the  $\omega\ll U$} limit. In this case the quantum states should follows 
\begin{equation}
\label{LL1Une0}
\varepsilon _{n,\mu \pm} \simeq \mu U+ \frac{\mu}{2U}n(n\pm1)\,\omega_o^2B^2, 
\end{equation}

\noindent clearly, in addition to the aperture of an energy gap of $2U$ between the spin-degenerate electrons/hole LLs, the presence of the gate voltage induces a deviation from the linear dependence in $B$ occurring at $U=0$, to a parabolic behavior with  $B$ instead. This is also a known result in the literature. \cite{Koshino,Vasilopoulus}. \\

\noindent ({\it ii}) {\it Weak Rashba-SOI in the regime } $U\gg\omega\gg\Gamma$. Here the LLs still behave quadratic in $B$ but with a spin-dependent shift linear in $B$ given as  

\begin{equation}
\label{LL2Une0}
\varepsilon _{n,\mu\pm }
\simeq 
\mu U +  \frac{\mu\, n(n\pm 1 )}{2U}\left(\omega_o^2B^2\pm \frac{\Gamma_o^2 B}{2}  \right ),
\end{equation}

\noindent the term proportional to $\Gamma_o^2$ is responsible for the anticrossings of the fan spectrum of the $\nu=\pm$ states.\\

\noindent ({\it iii}) {\it Strong Rashba-SOI in the regime } $U\gg\Gamma \gg \omega$. In contrast with the case of $U=0$,  at very large Rashba-SOI strengths the LL spectrum  follows (to leading order) a  linear behavior with $B$ for the positive chirality states $(\nu= +)$, whereas for the negative $(\nu= -)$ quantum states develops a cubic dependence instead. Explicitly they are given by, 
\begin{equation}
\label{LL3Une0}
\varepsilon _{n,\mu+ }
\simeq 
\mu U +  \frac{\mu\, \Gamma_o^2 }{2U}nB , \quad\quad  \varepsilon _{n,\mu- }
\simeq 
\mu   U + \frac{\mu\,\omega_o^4}{2U\Gamma_o^2}n(n^2-1)B^3.
\end{equation}

\noindent such drastic change on the field dependence of the LLs with different spin chirality states $(\nu= \pm)$ will enhance dramatically the degree of their spin-splitting and on the multiplicity of the level crossing as studied in the next section.

All of this discussed above holds for $n>2$. The cases $n=0,1$ are treated separately as in the condition for $U=0$.  In the limit $U\gg \omega\gg \Gamma$ (small Rashba SOI or large fields) $\varepsilon_{1\mu+}\simeq \mu(U+ \omega_o^2B^2/U)$, and $\varepsilon_{1\mu+}\simeq \mu(U + \Gamma_o^2B/U)$ for the large Rashba-SOI (or small Rashba-SOI) case ($U\gg \Gamma\gg\omega)$, whiles $\varepsilon_{1-}=-U$ for both regimes.  The case for the zero Landau level index is $\varepsilon_{0-}=U$, which also holds in both regimes.

\begin{figure}
\vspace{-0.3in} 
\hspace*{4.0cm}\includegraphics[width=4.0in,height=4.0 in]{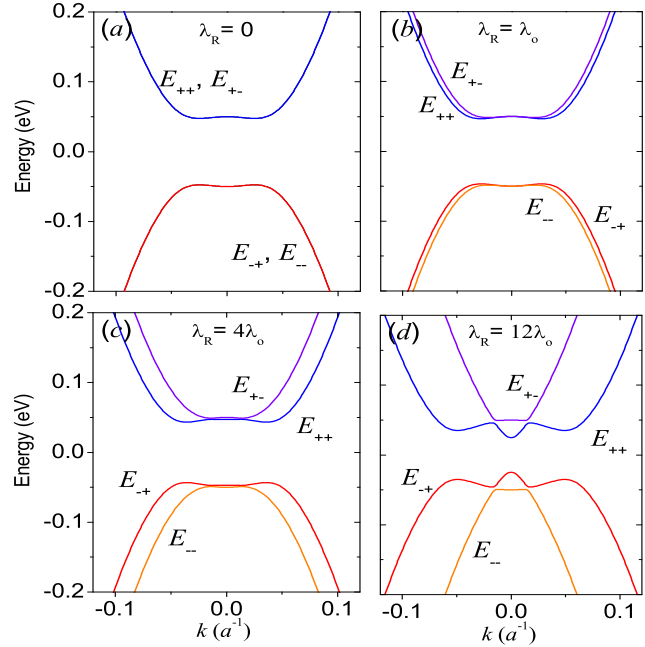} 
\vspace{-0.4in}
\caption{  (Color online) Low energy spectrum for biased bilayer graphene with Rashba-SOI. Here $U=0.050$ eV. The bias voltage induced gap decreases as the Rashba parameter $\lambda_R$ increase.} 
\label{Dispersion-Bias}
\end{figure}

\subsection{Spin-polarization of the Landau levels}
The eigenfunctions of Eq. (\ref{HLandau}) for the $n-$th Landau level of a given electron(hole) band $\mu$ in biased BLG are given in Appendix C. 
From  Eqs.\,(\ref{LL1}) and the orthogonality of the oscillator wave functions $\xi_n$, it follows that, the valley $\langle { \bn \tau^{\scriptscriptstyle(n)} }  \rangle_{\mu \nu}$ and $\langle  {\bn S}^{\scriptscriptstyle(n)}  \rangle_{\mu\nu}$ spin polarization lying in the plane of BLG   vanishes identically for all Landau levels.  General expressions for the valley and spin-polarization are provided in Eqs. (\ref{TauLL}) and (\ref{SLL}). We notice that the valley polarization in the perpendicular direction turns to be $k-$independent, and that for the unbiased  case, it vanishes for all LL. Furthermore, the $z$-th component of the spin polarization of the $n-$th LL with $U=0$, reduces to

\begin{eqnarray} 
\label{SznU0}
\langle { S_z^{\scriptscriptstyle(n)}} \rangle_{\mu\nu}
& =  & -  \frac{\nu}{2}\left( \cos\vartheta_{n\scriptscriptstyle -} +\cos \vartheta_{n\scriptscriptstyle +} \right)\nonumber  \\
& =  & -\frac{2\nu\, \omega^2}{\sqrt{4\omega^2+4n\,\omega^2\Gamma^2+\Gamma^4}}, 
\end{eqnarray}   

\noindent where, as before $\nu=\pm$ denotes the plus/minus $n-$LL of a given $\mu=\pm$ electron/hole branch. In the limit of high field $\langle { S_z^{\scriptscriptstyle(n)}} \rangle_{\mu\pm}\rightarrow \mp 1$, a full polarization is reached, and the states $\nu=\pm$ coincides with the spin-magnetization signs ($\mp$) of the LLs. If the limit $B\rightarrow 0$ is taken, then the spin-polarization $\langle { S_z^{\scriptscriptstyle(n)}} \rangle_{\mu\pm}\rightarrow \pm2(\omega_o^2/\Gamma_o^2)B$.

\section{Band structure properties in BLG with Rashba SOI} 

The low quasiparticle energy band structure for unbiased bilayer graphene with Rashba-SOI, as predicted by Eq.(\ref{EUnbias}) at zero field, is illustrated in Fig.1   for different values of $\lambda_R$ strength. In the absence of Rashba-SOI ($\lambda_R=0$) the well recognized parabolic spin-degenerated conduction and valence bands touching at its extrema at $k=0$ are depicted in Fig.1(a). For non-zero $\lambda_R$ the spin-degeneracy of the bands is broken inducing a $k$-linear spin-splitting of width $\Delta_s(k) =|E_{\scriptscriptstyle \pm,\mp}-E_{\scriptscriptstyle \pm,\pm}| = \alpha k$. For relatively weak Rashba-SOI ($4\beta^2k^2\gg \alpha^2$) the band dispersion follows a parabolic behavior, $\varepsilon^o _{\mu s}(k)\simeq \mu (\beta^2k^2 -{\scriptscriptstyle \frac{1}{2}}s\alpha k)$, as shown in Fig.1(b). On the other hand, if the condition $4\beta^2k^2\ll \alpha^2$ holds, then the relation (\ref{EUnbias}) evolves to a linear spectrum for the inner bands ($\varepsilon^o _{\mu s}(k)\simeq \mu \alpha k$ for $s=-$), and to a $k$-cubic spectrum for the outermost bands ($\varepsilon^o _{\mu s}(k)\simeq \mu \beta^2 k^3/\alpha$ for $s=+$) as plotted in Fig.1(d). Similar drastic changes in the bandstructure due to large Rashba-SOI strengths were reported earlier numerically by van Gelderen {\it et al.}\cite{vanGelderen} within a tight-binding framework; see for instance the low energy bands near the Dirac points of Fig.1(d) of that reference. 

\begin{figure} 
\hspace*{4.0cm}\includegraphics[width=4.3in,height=3.8 in]{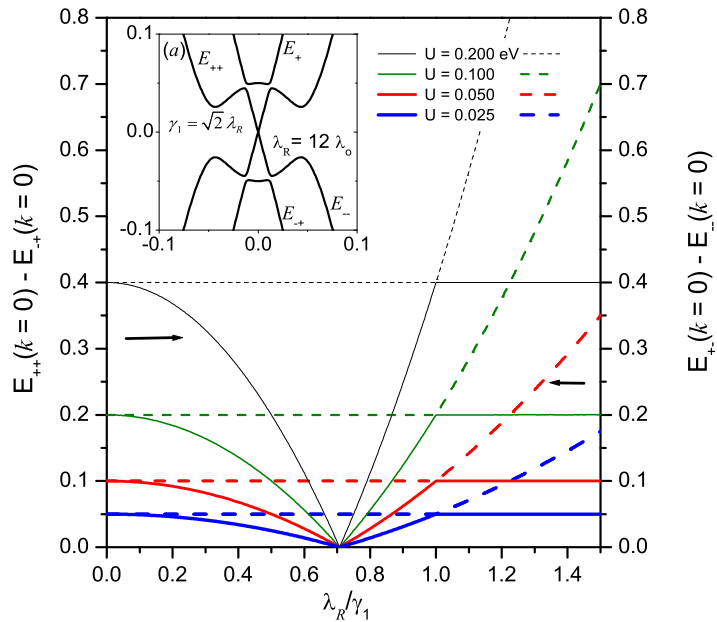}
\vspace{-0.4in}
\caption{(Color online) The Rashba-SOI effectively modulates the gap size of biased BLG. Inset ($a$)  shows the spectrum of the bands for the critical Rashba-SOI parameter that ensures a closing gap, $\lambda_R=\gamma_1/\sqrt{2}$ with $U=0.025$ eV, $\gamma_1=0.22$ eV and $\lambda_R=12\lambda_o$.  }
\label{Gap}
\end{figure}

At the intermediate regime (Fig.1(c)), the innermost bands interpolates from a $k$-linear behavior for electron/hole momentum very close to the Dirac point, to a $k$-cubic dependence for high momentum. In contrast, the $s=+$ bands seems to be well described by the cubic spectrum for all values of $k$. Such remarkable asymmetry behavior of the spectrum of BLG with Rashba-SOI is certainly unique, since it is not seen in monolayer graphene neither in semiconductors with the Rashba-type of SOI. These peculiar characteristics of the spectrum of BLG would have interesting consequences on the electronic and spin-transport properties.  

In Fig.2 we show the bilayer graphene low energy spectrum for finite bias voltage ($U=0.050$ eV) at various Rashba coupling strengths ($\lambda_R=0,\lambda_o,4\lambda_o$ and $12\lambda_o$). As in the unbias case, without Rashba-SOI, the bands are spin degenerated (Fig.2(a)). A gap of $2|U|$ at $k=0$ is opened between the conduction and valence bands turning BLG into a semiconductor. Moreover a band-bending appears at small wave-numbers (low momentum) due interplay with the bias gate voltage $U$. This is the so-called "Mexican-hat-like" shape of the lowest energy bands well reported in the literature. From Eq.(\ref{Eko}),  in this regime ($\lambda_R=0$) and $ka\lesssim 1$ the bands are reasonably well described by $\varepsilon_{\mu s}(k)\simeq \mu (U-\xi k^2+(\beta^2/2U)k^4)$. For nonzero $\lambda_R$ (Figs.2(b)-(d)) the spin-degeneracy is lifted. As the Rashba-SOI parameter increase the lowest/highest conduction/valence bands becomes more warped and the gap tend to decrease as the lowest conduction band $E_{\scriptscriptstyle ++}$ evolves from a Mexican-hat like shape to an inverted one, and vice versa for the highest valence band $E_{\scriptscriptstyle -+}$, see Fig.3(d) calculated using Eq.(\ref{Ek}).  The behavior of the gaps $\Delta_{g+}=E_{\scriptscriptstyle ++}-E_{\scriptscriptstyle -+}$ (for the innermost bands) and $\Delta_{g-}=E_{\scriptscriptstyle +-}-E_{\scriptscriptstyle --}$ (for the outmost bands) as a function of the ratio $\lambda_R/\gamma_1$ is plotted in Fig.3 for different bias voltages $U$. The gap $\Delta_{g+}$ in BLG closes as the Rashba parameter increases, reaching its minimum (zero gap) at $\lambda_R =\gamma_1/\sqrt{2}$, to then gradually and linearly opens again as $\lambda_R/\gamma_1$ is increase up to $1$. For $\lambda_R/\gamma_1>1$ the gap $\Delta_{g+}$ remains constant. Inset ($a$) of Fig.3 shows the bandstructure for biased BLG with $U=0.025$ eV illustrating the zero gap condition. An analogous behavior can be seen in the numerical plot depicted in Fig.7(b) of Ref.\cite{vanGelderen}. Notice that our analytical low effective modeling for the bilayer graphene bandstructure allow us to capture also these anomalous behavior of the lowest bands (near the Dirac points) under finite bias and relatively large Rashba coupling. In addition it  predicts that the closing of the gap occurs provided that $\lambda_R=\gamma_1/\sqrt{2}$,  regardless of the magnitude of the bias voltage applied.  

\begin{figure} 
\vspace{-0.0cm}
\hspace*{4.0cm}\includegraphics[width=4.0in,height=3.3 in]{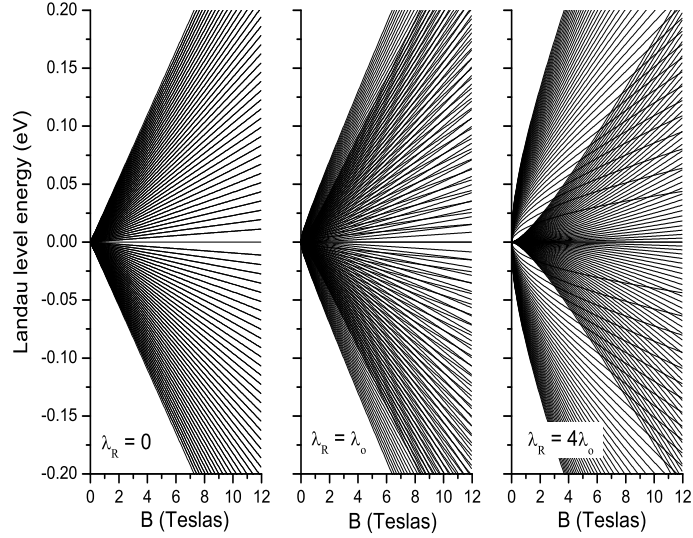}
\vspace{-0.4in}
\caption{ Spectrum of the LLs for bilayer graphene with Rashba-SOI within the low effective low energy theory. A rather unusual characteristics of the LLs is predicted to occur at large values of the Rashba-SOI strength.  }
\label{ LL-Unbias}
\end{figure}

\section{Landau level spectrum in BLG with Rashba-SOI}
The Landau level energy spectrum as a function of magnetic field (up to 12\,T) for BLG with Rashba-SOI according to formula Eq.(\ref{LL-BLG}) is plotted (from $n=0$ to $n=41$) in Fig.4 and Fig.5 for the unbiased and biased case, respectively. The zero gate voltage ($U=0$) and without Rashba-SOI case shows a LL fan diagram which is linear with $B$ and degenerate in spin (Fig.4(a)), as expected, because of the parabolic behavior of the energy bands and since there is no spin-dependent mechanism here to break the spin-symmetry. 
When the Rashba-SOI is present the spin-degeneracy is lifted inducing multiple crossings of the LLs at the Fermi energy, similar as it occurs in semiconductor 2DEGs with Rashba SOI, and stems because for sizable $\lambda_R$ strengths the LLs with high index and with spin chirality $s=+$ have lower energies than those with spin chirality $s=-$ as the field is increased. This is basically an intermediate regime between those LLs discussed in section {\bf 3.1}({\it ii}) and {\bf 3.1}({\it iii}). Notice that for a relatively weak intensity of the Rashba parameter ($\lambda_R=\lambda_o$) the LLs behave roughly linear with the field (Fig.4(b)), no matter its spin chirality.  

However, for large values ($\lambda_R=4\lambda_o$) a drastic and unusual change in the LLs spectrum arises (Fig.4(c)); the LLs with spin $\nu=+$ evolves as $B^{1/2}$ whiles for those with  $\nu=-$ develops   a  $B^{3/2}$ dependence as described in section {\bf 3.1}({\it iii}). Such difference in the field dependence effectively squeezes the LLs with $\nu=-$ to lower energies as $\lambda_R$ is increased promoting multiple crossings between the LLs. This surprising result suggests a strong spin polarization of the Landau levels induced by a significant increase of the Rashba parameter.  

\begin{figure} 
\hspace*{4.0cm}\includegraphics[width=4.0in,height=3.3 in]{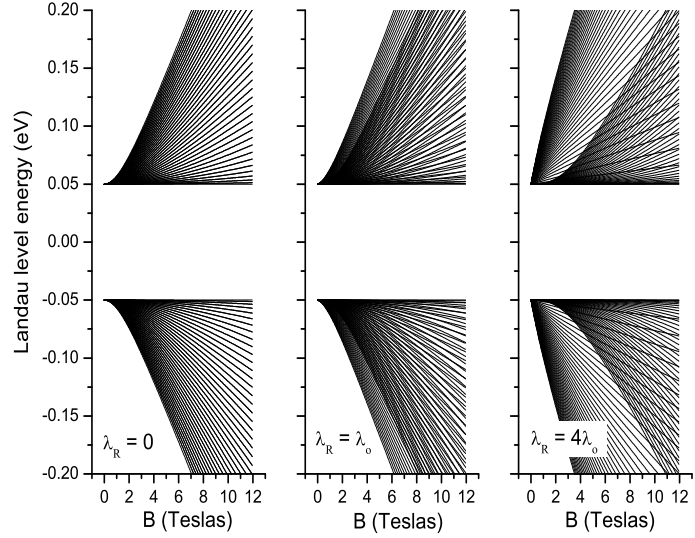}
\vspace{-0.4in}
\caption{The presence of a bias voltage split in two the fan diagram opening a gap of $2U$ between the electron and hole LLs.   }
\label{LLbias}
\end{figure}

The LLs for the biased $U\ne0$ and without Rashba coupling case shows instead a parabolic behavior with $B$, and the fan diagram is split by a gap of $2U$ (Fig.5(a)) as also described by Eq.(\ref{LL1Une0}). The linear behavior observed for $U=0$ and $\lambda_R=\lambda_o$ of the LLs transforms as well to a $B^2$ dependence for $U\ne0$, as predicted also by Eq.(\ref{LL2Une0}). The presence of the Rashba-SOI manifest as well as crossings of the LLs as seen in Fig.5(b).  The behavior illustrated in Fig.5(c) at relatively large $\lambda_R$ is similar to that seen in Fig.4(c) but with an open gap and with almost a linear behavior with field for $\nu=+$, and roughly a cubic dependence in {\it B} for the $\nu=-$ LLs, se also Eq.(\ref{LL3Une0}). Such asymmetric response at large Rashba-SOI in BLG contrast radically with the linear behavior with field that occurs in single layer graphene\cite{Rashba} at the same regime.

\section{Summary}
We have studied the problem of the influence of the Rashba spin-orbit 
interaction on the bandstructure of biased and unbiased bilayer graphene. 
Using low energy effective theory we have derived a low energy Hamiltonian 
for bilayer graphene in the presence of an external magnetic field and 
spin-orbit interactions. Analytical formulae for the energy spectrum of a 
graphene bilayer with Rashba spin-orbit interaction are obtained.
We show that for relatively weak Rashba coupling the spin-degeneracy of the 
electron and hole bands is broken inducing a $k$-linear spin-splitting, very 
similar to that  found in semiconductors heterostructures. At the intermediate 
strengths of the Rashba effect, the innermost bands interpolates from a 
$k$-linear behavior at small momentum, to a $k$-cubic dependence at high momentum. 
In contrast, the outermost bands seems to be well described by the cubic 
spectrum for all values of $k$. For large values of the Rashba coupling there 
is a remarkable warping behavior of the spectrum near the Dirac point. Such 
behavior  is unique in biased bilayer graphene. It is found that the 
bias-induced gap in bilayer graphene decreases as the Rashba is increased, 
showing a behavior resembling  a topological insulator transition phenomena. 
These peculiar characteristics of the spectrum of bilayer graphene with 
Rashba spin-orbit interaction may have important consequences on its 
electronic and spin-transport properties.  

We also obtained an analytical expression for the Landau levels and 
spin-polarization in biased bilayer graphene with Rashba effect valid 
in the low bias regime. It is further predicted the appearance of an 
unexpected asymmetric spin-splitting and crossings of the Landau levels  
due the combined effect between the Rashba interaction and the bias voltage. 
These results suggests significant consequences  on    the 
Shubnikov-de Hass oscillations and magnetotransport in bilayer graphene as 
quantum and spin Hall effects, under the presence of sizable Rashba 
spin-orbit interaction in the range of few meVs.

\ack
F.M. is thankful to J. Hausinger for useful discussions. This work was supported by Deutsche Forschungsgemeinschaft via GRK 1570, and by the Mexican grant 
Papiit-UNAM No. IN109911-3.

\appendix
\section{Derivation of the low energy Hamiltonian}
In this appendix we derive the low-energy Hamiltonian for bilayer graphene in 
the presence of both, intrinsic and Rashba spin-orbit interaction, as well as 
Zeeman effect. The low energy Hamiltonian is obtained via L\"owdin partitioning, 
also known as van Vleck's perturbation theory in the context of atomic 
physics.\cite{van-Vleck,Cohen,Zhang} First, it is convenient to express 
the Hamiltonian $H_k$ of Eq.(5) in the new spin-dependent atomic basis
$|\tilde \Psi^{\dagger}_{K}\rangle =\{\psi_{A_{2\uparrow}}, \psi_{A_{2\downarrow}},\psi_{B_{1\uparrow}}, \psi_{B_{1\downarrow}},\psi_{A_{1\uparrow}}, \psi_{A_{1\downarrow}},\psi_{B_{2\uparrow}}, \psi_{B_{2\downarrow}} \}$  leading to the $8\times8$ Hamiltonian

\begin{equation}
\fl
\scriptsize
H_k=\left(
\begin{array}{cccccccc}
 -U+\eta +\Delta  & 0 & 0 & 0 & 0 & 0 & \gamma  \pi   & 0 \\
 0 & -U-\eta -\Delta  & 0 & 0 & 0 & 0 & i \lambda_{R}  & \gamma  \pi   \\
 0 & 0 & U-\eta +\Delta  & 0 & \gamma  \pi^{\dagger}& -i \lambda_{R}  & 0 & 0 \\
 0 & 0 & 0 & U+\eta -\Delta  & 0 & \gamma  \pi^{\dagger} & 0 & 0 \\
 0 & 0 & \gamma  \pi  & 0 & U+\eta +\Delta  & 0 & \gamma _1 & 0 \\
 0 & 0 & i \lambda_{R}  & \gamma  \pi   & 0 & U-\eta -\Delta  & 0 & \gamma _1 \\
 \gamma  \pi^{\dagger} & -i \lambda_{R}  & 0 & 0 & \gamma _1 & 0 & -U-\eta +\Delta  & 0 \\
 0 & \gamma  \pi^{\dagger} & 0 & 0 & 0 & \gamma _1 & 0 & -U+\eta -\Delta 
\end{array}
\right), 
\end{equation}

\noindent which now can be written as the sum $H_k={\cal H}_o+W$, with  $\{U,\gamma _1,\Delta,\eta  \}\in{\cal H}_o$ and $\{\lambda_{R},\gamma\pi,\gamma  \pi^{\dagger},  \}\in W$ where

\begin{equation}
\fl
\footnotesize
{\cal H}_o=\left(
\begin{array}{cc}
 {\cal H}_{+}  & 0 \\
 0 & {\cal H}_{-}
\end{array}
\right),\quad
{\cal H}_{\pm} =\left(
\begin{array}{cccc}
 \mp U+\Delta+\eta   & 0 & \gamma_1\,\delta_{\pm} & 0 \\
 0 & \mp U-\Delta-\eta   & 0 & \gamma_1\,\delta_{\pm} \\
 \gamma_1\,\delta_{\pm} & 0 & \pm U+\Delta-\eta  & 0 \\
 0 & \gamma_1\,\delta_{\pm} & 0 & \pm U-\Delta-\eta 
\end{array}
\right), 
\end{equation}

\noindent with  $\delta_{+}=0$ , and $\delta_{-}=1$. On the other hand

\begin{equation}
\footnotesize 
 W=\left(
\begin{array}{cc}
 0 & H_s \\
 H_s & 0
\end{array}
\right), \quad \mbox {with} \quad
H_s=\left(
\begin{array}{cccc}
 0 & 0 & \gamma  \pi & 0 \\
 0 & 0 & i \lambda_R  & \gamma  \pi \\
 \gamma  \pi^{\dagger} & -i \lambda_R  & 0 & 0 \\
 0 & \gamma  \pi^{\dagger} & 0 & 0
\end{array}
\right)
\end{equation}

Notice that $H_s$ is nothing but the single-layer graphene Hamiltonian with Rashba-SOI  in the basis $\{\psi_{A_{2(1)\uparrow}}, \psi_{A_{2(1)\downarrow}},\psi_{B_{1(2)\uparrow}}, \psi_{B_{1(2)\downarrow}} \}$. 

Bilayer graphene with RSOI, ISOI and Zeeman effect has (in general) eight non-degenerated levels. The levels are given by the eigenvalues of the Hamiltonian ${\cal H}_o$,
 
\begin{eqnarray}
\footnotesize
E_1^o & = & -\Delta -\sqrt{\gamma _1^2+(U+\eta )^2 }\\  
E_2^o & = & \Delta -\sqrt{\gamma _1^2+(U-\eta )^2} \\
E_3^o & = & -U-\eta -\Delta \\
E_4^o & = & -U+\eta +\Delta \\
E_5^o & = & U-\eta +\Delta \\
E_6^o & = & U+\eta -\Delta \\
E_7^o & = & -\Delta +\sqrt{\gamma _1^2+(U-\eta )^2}\\
E_8^o & = & \Delta +\sqrt{\gamma _1^2+(U+\eta )^2}
\end{eqnarray}

The eigenvectors $|\Psi_\mu\rangle$   of ${\cal H}_o$ (with $\mu=1,8$) can be written as a column vectors of the matrix

\begin{equation} 
\left(
\begin{array}{cccccccc}
 0 & 0 & 0 & 1 & 0 & 0 & 0 & 0 \\
 0 & 0 & 1 & 0 & 0 & 0 & 0 & 0 \\
 0 & 0 & 0 & 0 & 1 & 0 & 0 & 0 \\
 0 & 0 & 0 & 0 & 0 & 1 & 0 & 0 \\
 0 & -\cal{S}_+ & 0 & 0 & 0 & 0 & 0 & \cal{C}_+ \\
 -\cal{S}_- & 0 & 0 & 0 & 0 & 0 & \cal{C}_- & 0 \\
 0 & \cal{C}_+ & 0 & 0 & 0 & 0 & 0 & \cal{S}_+ \\
 \cal{C}_-  & 0 & 0 & 0 & 0 & 0 & \cal{S}_- & 0
\end{array}
\right)
\end{equation}

\noindent with ${ \cal C}_\pm= \cos \left(\vartheta_\pm /2\right)$,  ${\cal S}_\pm=  \sin \left(\left.\vartheta _\pm\right/2\right)$ and $\tan \,\vartheta_\pm=\gamma_1/(U\pm \eta)$. Because of the strength of the parameter $\gamma_1$, the energy levels of the subspace   of high energy,  $  {\cal E}^o_{i\mathit{a}}\in \left\{E_1^o,E_2^o,E_7^o,E_8^o\right\}$, and the energy levels of the subspace with low energy,   $ {\cal E}^o_{j\mathit{b}} \in \left\{E_3^o,E_4^o,E_5^o,E_6^o\right\} $  are  well separated from each other ({\it i.e.} $|{\cal E}^o_{i\mathit{a}}-{\cal E}^o_{j\mathit{a}}|\sim |{\cal E}^o_{i\mathit{b}}-{\cal E}^o_{j\mathit{b}}| \ll |{\cal E}^o_{i\mathit{a}}-{\cal E}^o_{j\mathit{b}}| \sim |\gamma_1|$ ). The low energy Hamiltonian for bilayer graphene can thus be obtained through the unitary transformation $ {\cal H}=e^{i S}H_k e^{-i S}$, in which the $S$ matrix elements are given by
 
\begin{eqnarray} 
S_{\mu \nu } & = &\frac{i W_{\mu \nu }}{E_{\nu }^o-E_{\mu }^o}+i\sum _{\mu '} \frac{W_{\mu \mu '}W_{\mu '\nu }}{\left(E_{\nu }^o-E_{\mu }^o\right)\left(E_{\nu }^o-E_{\mu '}^o\right)}\nonumber \\ 
             & - & i\sum _{\nu '} \frac{W_{\mu \nu '}W_{\nu '\nu }}{\left(E_{\nu }^o-E_{\mu }^o\right)\left(E_{\nu '}^o-E_{\nu }^o\right)}, \nonumber
\end{eqnarray}

\noindent with $S_{\mu \nu }=\left(S_{\nu \mu }\right){}^{\dagger}$   and $ W_{\mu \nu }=<\Psi _{\mu }|W|\Psi _{\nu }>.$ The low energy matrix elements of the effective Hamiltonian up to second order in $1/\gamma_1$ are determined by
\begin{eqnarray} 
 {\cal H}_{\mu \mu '} & = & E_{\mu }^o\delta _{\mu \mu '}+W_{\mu \mu '} \nonumber \\  
                     & + & \frac{1}{2}\sum _{\nu } W_{\mu \nu }W_{\nu \mu '}\left(\frac{1}{E_{\mu }^o-E_{\nu }^o}+\frac{1}{E_{\mu '}^o-E_{\nu }^o}\right)+ {\cal O}(2), \nonumber  
\end{eqnarray}

\noindent with ${\cal H}_{\mu \nu }=\left({\cal H}_{\nu \mu }\right){}^{\dagger}$ and  $\mu,\mu ' \in  \{3,4,5,6\}$ and $\nu,\nu ' \in  =\{1,2,7,8\}$. The effective Hamiltonian matrix elements reads, 

\begin{eqnarray} 
{\cal H}_{33} & = & -U-\eta -\Delta +\frac{2(U+\eta +\Delta )\lambda _R^2}{\gamma_1^2}+\frac{2\gamma^2 U}{\gamma_1^2}\,\pi^{\dagger} \pi \nonumber   \\
{\cal H}_{34} & = & \frac{i(2U+\eta +\Delta )\lambda _R}{\gamma _1{}^2}\pi \nonumber\\
{\cal H}_{35} & = & \frac{-2i \lambda _R\gamma}{ \gamma_1^2}\,\pi^{\dagger} \nonumber\\
{\cal H}_{36} & = & -\frac{\gamma ^2}{\gamma _1}\,\left(\pi^{\dagger}\right)^2 \nonumber\\
{\cal H}_{44} & = & -U+\eta +\Delta +\frac{2U \gamma ^2}{\gamma _1{}^2}\,\pi^{\dagger}\pi \nonumber\\
{\cal H}_{45} & = & H_{36} \nonumber\\
{\cal H}_{46} & = & 0 \nonumber\\
{\cal H}_{55} & = & U-\eta +\Delta -\frac{2(U-\eta -\Delta )\lambda _R{}^2}{\gamma _1{}^2}-\frac{2U \gamma ^2}{\gamma _1{}^2}\,\pi\pi^{\dagger} \nonumber \\
{\cal H}_{56} & = & \frac{2i \lambda _R\gamma }{\gamma _1{}^2}(U+\Delta )\,\pi^{\dagger} \nonumber\\
{\cal H}_{66}  & = & U+\eta -\Delta -\frac{2U \gamma ^2}{\gamma _1{}^2}\,\pi\pi^{\dagger}. \nonumber  
\end{eqnarray}

These matrix elements form  the desired $4\times4$  low energy Hamiltonian for bilayer graphene  

\begin{equation}
\label{Himplicit}
\mathcal{H}=\left(
\begin{array}{cccc}
 \mathcal{H}_{33} & \mathcal{H}_{34} & \mathcal{H}_{35} & \mathcal{H}_{36} \\
 \mathcal{H}_{34}^{\dagger} & \mathcal{H}_{44} & \mathcal{H}_{45} & \mathcal{H}_{46} \\
 \mathcal{H}_{35}^{\dagger}  & \mathcal{H}_{45}^{\dagger} & \mathcal{H}_{55} & \mathcal{H}_{56} \\
 \mathcal{H}_{36}^{\dagger} & \mathcal{H}_{46}^{\dagger} & \mathcal{H}_{56}^{\dagger} & \mathcal{H}_{66}
\end{array}
\right),
\end{equation}

\noindent which written in terms of suitable Kronecker products leads to Eq. (\ref{TotalH}). Note that Eq. (\ref{Himplicit}) was projected on the basis set $\{\psi_{A_{2\downarrow}}, \psi_{A_{2\uparrow}},\psi_{B_{1\uparrow}}, \psi_{B_{1\downarrow}}\}$.  
 
{\it Energy dispersion.} Consider vanishing intrinsic SOI and no magnetic field ($\eta=\Delta=0$). To a good approximation we can neglect the off-diagonal terms that goes as $1/\gamma_1^2$ in the effective Hamiltonian. In such a case the eigenvalues of (\ref{Himplicit}) are readily obtained

\begin{equation} 
\label{Ek}
E_{\lambda s}(k)=\frac{\lambda }{\sqrt{2}}\sqrt{ U^2+(U-\rho )^2+{\cal A}\,k^2+2{\cal B}\, k^4-s\sqrt{\Upsilon }},
\end{equation}
\noindent here ${\lambda  }$  indicates the electron $(+)$ and hole $(-)$ branches, whereas $s=\pm$ labels the spin state chirality, ${\cal A}=\Lambda-4\tilde\xi \,U$, ${\cal B}=\tilde \beta^2+\tilde\xi^2$, $\Lambda =\tilde\alpha ^2+2 \tilde\xi  \rho$ and  $\Upsilon =4 \alpha ^2 \tilde\beta ^2k^6 +\left(\Lambda^2+4\rho ^2\tilde \beta^2\right)k^4-2 \rho \Lambda (2 U-\rho )k^2+\rho^2 (2 U-\rho )^2$, where we have introduced the parameters
\begin{equation} 
\tilde \alpha =\frac{2\hbar\gamma\,\lambda _R  }{\gamma _1},\,\,\,\tilde \beta =\frac{\hbar^2\gamma ^2}{\gamma _1},\,\,\,\rho =\frac{ 2U \lambda _R}{\gamma _1{}^2}, \,\,\,\tilde\xi =\frac{2U \hbar^2\gamma ^2}{\gamma _1^2}.
\end{equation} 

In the limit $\rho\rightarrow 0$ Eq. (\ref{Ek}) reduces to Eq.(14) with $\lambda=\mu$.

\section{Eigenvalues of the low energy Hamiltonian}
 
By squaring the low energy Hamiltonian (\ref{Heff}), a straightforward diagonalizable system is obtained at zero magnetic field ($B=0$).

\begin{equation} 
\fl
\scriptsize
\label{H2}
{\cal H}_k^2=\left(
\begin{array}{cccc}
 {(U-\xi k^2)}^2+\beta^2k^4 & -i \alpha  \beta \, k^2 \,k_- & 0 & 0 \\
 i \alpha  \beta \, k^2\, k_+ &{(U-\xi k^2)}^2+ \alpha ^2k^2+\beta ^2k^4 & 0 & 0 \\
 0 & 0 & {(U-\xi k^2)}^2+\alpha ^2k^2+\beta ^2k^4 & -i \alpha  \beta \, k^2\, k_- \\
 0 & 0 & i \alpha  \beta \, k^2\, k_+ & {(U-\xi k^2)}^2+\beta ^2k^4
\end{array}
\right)\, ,
\end{equation}

\noindent with $k_{\pm}=k_x \pm k_y$ and the eigensystem ${\cal H}_k^2|\chi \rangle =\varepsilon_k^2|\chi\rangle $ yields the eigenvalues
 
\begin{equation} 
\varepsilon _{k,\pm }^2 ={(U-\xi k^2)}^2+\frac{1}{2} k^2 \left(\alpha ^2+2   \beta ^2 k^2\pm  \alpha  \sqrt{\alpha ^2+4  \beta ^2k^2}\right), 
\end{equation} 
\noindent whereas its eigenvectors $|\chi_j \rangle$ arranged as column vectors form a unitary matrix    
 
\begin{equation} 
\mathbb{U} =\left(
\begin{array}{cccc}
 -i\,  {\sin}\left( \theta /2 \right) & 0 & i\,  {\cos}\left(\theta /2\right ) & 0 \\
   e^{i \phi } {\cos}\left(\theta /2\right ) & 0 & e^{i \phi } {\sin}\left(\theta /2\right )  & 0 \\
 0 &  -i \, {\cos}\left(\theta /2\right) & 0 & i\,  {\sin}\left(\theta /2\right) \\
 0 & e^{i \phi } {\sin}\left(\theta /2\right)  & 0 &  e^{i \phi } {\cos}\left(\theta /2\right)  
\end{array}
\right),
\end{equation}

\noindent with  $\tan  \theta  = 2\beta  k/\alpha$, $ e^{i \phi }=(k_x+i k_y)/k$ and $\mathbb{U}^{\dagger}={\mathbb U}^{-1}$. 
In the basis of the  eigenvectors of ${\cal H}_k^2$,  the $4\times4$ Hamiltonian ${\cal H}_k$ is conveniently transformed as follows 

\begin{equation} 
 {\tilde {\cal H}}_k= \mathbb{  U}^{\dagger}{\cal H}_k\mathbb{  U}=\left(
\begin{array}{cccc}
 -{U}+\xi k^2 & r_+^* & 0 & s^{*} \\
 r_+ &  \,{U}-\xi k^2 & -s & 0 \\
 0 & -s^{*}& -{U}+\xi k^2 & r_-^* \\
 s & 0 & r_- & \, {U}-\xi k^2
\end{array}
\right), 
\end{equation}
\noindent and $|  \tilde \psi_k \rangle=\mathbb{  U}^{-1}|\psi_k\rangle$, where
\begin{eqnarray} 
r_\pm    & = &     -\frac{1}{2}\,ke^{2i \phi }\left[2\beta k\, {\sin}\,\theta+\alpha ( {\cos}\,\theta  \mp 1)\right] \nonumber \\
         & = & -\frac{1}{2}\,ke^{2i \phi }\left[\sqrt{\alpha^2+4\beta^2k^2}\pm \alpha \right] 
\end{eqnarray}
\noindent with 
\begin{equation}
 {\sin}\,\theta=\frac{2\beta k}{\sqrt{\alpha^2+4\beta^2k^2}}, \quad  {\cos}\,\theta=\frac{\alpha}{\sqrt{\alpha^2+4\beta^2k^2}} 
\end{equation}

\noindent whereas
\begin{equation}
s      =       -\frac{1}{2}ke^{2i \phi }[2\beta k\, {\cos}\,\theta-\alpha\,    {\sin}\,\theta   ]=0.
\end{equation}

The eigenvalues of ${\tilde {\cal H}}_k$ are determined from $\varepsilon _{\mu s}(k)=\mu \sqrt{(U-\xi k^2)^2+\left|r_{s}\right|^2 }$, with $\mu=\pm$ for the electron/hole band and $s=\pm$ labeling the  spin chirality state.   Explicitly

\begin{equation}
\label{Eko}
\varepsilon _{\mu s}(k)=\frac{\mu}{2} \sqrt{4\, (U-\xi k^2)^2+k^2\left(\sqrt{\alpha^2+4 k^2 \beta^2}-s\, \alpha \right)^2 }
\end{equation}

\noindent  The eigenvectors of $\tilde {\cal H}_k$ are given by
 
\begin{equation} 
\fl
|\tilde\psi _1\rangle=\frac{1}{\sqrt{1+{(\chi}_{ \scriptscriptstyle +}^{\scriptscriptstyle{(-)}})^2}} \left(
\begin{array}{c}
 i\,e^{-2i \phi }\chi_{ \scriptscriptstyle +}^{\scriptscriptstyle{(-)}}  \\
 1 \\
 0 \\
 0
\end{array}
\right), \quad 
|\tilde\psi _2\rangle=\frac{1}{\sqrt{1+{(\chi}_{ \scriptscriptstyle -}^{\scriptscriptstyle{(-)}})^2}} \left(
\begin{array}{c}
 0 \\
 0 \\
 i\,e^{-2i \phi }\chi_{ \scriptscriptstyle -}^{\scriptscriptstyle{(-)}} \\
 1
\end{array}
\right), 
\end{equation}
  
\begin{equation}
\fl
|\tilde\psi _3\rangle=\frac{1}{\sqrt{1+{(\chi}_{ \scriptscriptstyle +}^{\scriptscriptstyle{(+)}})^2}}\left(
\begin{array}{c}
 i\,e^{-2i \phi }{\chi}_{ \scriptscriptstyle +}^{\scriptscriptstyle{(+)}} \\
 1 \\
 0 \\
 0
\end{array}
\right),\quad 
 |\tilde\psi _4\rangle=\frac{1}{\sqrt{1+{(\chi}_{ \scriptscriptstyle -}^{\scriptscriptstyle{(+)}})^2}}\left(
\begin{array}{c}
 0 \\
 0 \\
 i\,e^{-2i \phi }{\chi}_{ \scriptscriptstyle -}^{\scriptscriptstyle{(+)}}\\
 1
\end{array}
\right)
\end{equation}
 
 The eigenvectors of ${\cal H}_k$ are finally given by $|   \psi_k \rangle=\mathbb{  U}  |\tilde\psi_k\rangle$ leading to Eqs.(\ref{eigen1}) and (\ref{eigen2}) with ${\chi}_{s}^{\scriptscriptstyle{(\mu)}}$ as given by Eq.(\ref{R}).
 
\section{Landau levels in BLG with Rashba coupling}
 
Here we  outline the derivation of the Landau levels in biased bilayer graphene with Rashba SOI. We follow the same approach used in Appendix B. Squaring the Hamiltonian (\ref{HLandau}) gives the block-diagonal matrix

\begin{equation} 
\fl
\scriptsize
\label{HL2}
{\cal H}_n^2=
 \left(
\begin{array}{cccc}
 {  U}^2+(n-1) n \omega ^2 & -i   n\sqrt{ n-1   }\, \Gamma   \omega  & 0 & 0 \\
 i  n  \sqrt{ n-1   }\, \Gamma   \omega  & {  U}^2+n \left(\Gamma ^2+(n+1)   \,\omega ^2\right) & 0 & 0 \\
 0 & 0 & {  U}^2+n \left(\Gamma ^2+(n-1)\, \omega ^2\right) & -i  n  \sqrt{  n+1} \,\Gamma\,  \omega  \\
 0 & 0 & i   n  \sqrt{n+1}\, \Gamma   \omega  & {  U}^2+n (n+1)\, \omega ^2
\end{array}
\right) ,
\end{equation}

\noindent the eigensystem ${\cal H}_n^2|\varphi_n \rangle =\varepsilon_n^2|\varphi_n\rangle $ leads to the eigenvalues

\begin{equation} 
\varepsilon _{n,\pm }^2 =\frac{1}{2} \left(2 \,{  U}^2+2 n^2 \omega ^2+n \Gamma ^2\pm n \sqrt{4 \omega ^4+4 n \omega ^2 \Gamma ^2+\Gamma ^4}\right), 
\end{equation} 
\noindent and its corresponding eigenvectors $|\varphi_{nj} \rangle$ written as column vectors form the   matrix,    
 
\begin{equation} 
\fl
{\mathbb V}=\left(
\begin{array}{cccc}
 0 & i \cos \vartheta_{n\scriptscriptstyle +}   & 0 & -i \sin \vartheta_{n\scriptscriptstyle +}   \\
 0 & \sin \vartheta_{n\scriptscriptstyle +}   & 0 & \cos \vartheta_{n\scriptscriptstyle +}  \\
 i \cos \vartheta_{n\scriptscriptstyle -}   & 0 & -i \sin \vartheta_{n\scriptscriptstyle -}   & 0 \\
  \sin \vartheta_{n\scriptscriptstyle -}   & 0 &   \cos \vartheta_{n\scriptscriptstyle -}   & 0
\end{array}
\right)  , \quad \tan  (\vartheta_{n\pm})  = \frac{2\sqrt{n\mp1}\,\omega\,  \Gamma}{2\,\omega^2\pm\Gamma^2}
\end{equation}

\noindent and ${\mathbb V}^{\dagger}{\mathbb V}=  1  $. 
In the basis of the  eigenvectors of ${\cal H}_n^2$,  the $4\times4$ Hamiltonian ${\cal H}_n$ in Eq.\,(\ref{HLandau}) is now transformed as follows 
 
\begin{equation} 
\fl
 {\tilde {\cal H}}_n= \mathbb{  V}^{\dagger}{\cal H}_n\mathbb{  V}=\left(
\begin{array}{cccc}
  {\cal U} & Q_{n-} & 0 & u_n  \\
 Q_{n-}  &  -{\cal U} & v_n & 0 \\
 0 & v_n&  {\cal U} & Q_{n+} \\
 u_n & 0 & Q_{n+}  & - {\cal U}
\end{array}
\right), \quad   {\rm and}   \quad  |  \tilde \phi_n \rangle=\mathbb{  V}^{-1}|\phi_n\rangle, 
\end{equation}
\noindent where
\begin{eqnarray} 
\label{Qn}
\fl
Q_{n\pm}         =   - \omega\left(\sqrt{n(n\pm1)}\cos \vartheta_{+} \cos \vartheta_{-}    +   \sqrt{n(n\mp1)}\sin \vartheta_{+}  \sin \vartheta_{-}\right)   \mp \sqrt{n}\,\Gamma \cos \vartheta_{\pm} \sin \vartheta_{\mp}  \nonumber \\
\label{Qnbis}
              =     - 4\omega^2 \Gamma\frac{\sqrt{n(n^2-1)\,{\cal N}_n}}{ \sqrt{  {\cal D}_{1\mp} {\cal D}_{2\pm } } } ,   \quad n>1
\end{eqnarray}

\noindent where we have used the useful relationships,

\begin{eqnarray}
\cos\vartheta_{n+}   =   \frac{2\omega^2+\Gamma^2 +\sqrt{{\cal N}_n}}{\sqrt{{\cal D}_{1+}}}  =\frac{2\sqrt{n-1}\,\omega\,\Gamma}{\sqrt{{\cal D}_{1-}}},  \\
 \sin\vartheta_{n+}   =  \frac{2\sqrt{n-1}\,\omega\,\Gamma}{\sqrt{{\cal D}_{1+}}}= -\frac{2\beta^2+\Gamma^2 -\sqrt{{\cal N}_n}}{\sqrt{{\cal D}_{1-}}}   \\
\cos\vartheta_{n-}  =   \frac{2\omega^2-\Gamma^2 +\sqrt{{\cal N}_n}}{\sqrt{{\cal D}_{2+}}}  =\frac{2\sqrt{n+1}\,\omega\,\Gamma}{\sqrt{{\cal D}_{2-}}}, \\
 \sin\vartheta_{n-}   =  \frac{2\sqrt{n+1}\,\omega\,\Gamma}{\sqrt{{\cal D}_{2+}}}= -\frac{2\beta^2-\Gamma^2 -\sqrt{{\cal N}_n}}{\sqrt{{\cal D}_{2-}}}.
\end{eqnarray}

\noindent with the definitions of the parameters
\begin{eqnarray}
 {\cal N}_n & =  &  4\,\omega^2(\omega^2+n\Gamma^2) + \Gamma^4, \label{Nn} \\ 
 {\cal D}_{1\pm} & = & (4n-1)\omega^2\Gamma^2+\left(2\omega^2+\Gamma^2\pm \sqrt{{\cal N}_n} \right)^2, \label{D1}\\
{\cal D}_{2\pm} & = & (4n+1)\omega^2\Gamma^2+\left(2\omega^2-\Gamma^2\pm \sqrt{{\cal N}_n} \right)^2. \label{D2}
\end{eqnarray}  

\noindent Notice that 
\begin{equation}
\fl
\footnotesize
u_n       =    - \omega\left(\sqrt{n(n+1)}\cos \vartheta_{n\scriptscriptstyle +} \sin \vartheta_{n\scriptscriptstyle-}    -   \sqrt{n(n-1)}\sin \vartheta_{n\scriptscriptstyle+}  \cos \vartheta_{n\scriptscriptstyle-}\right)   - \sqrt{n}\,\Gamma \cos \vartheta_{n\scriptscriptstyle+} \cos \vartheta_{n\scriptscriptstyle-}       =0,  
\end{equation}

\begin{equation}
\fl
v_n        =    - \omega\left(\sqrt{n(n+1)}\cos \vartheta_{n\scriptscriptstyle-} \sin \vartheta_{n\scriptscriptstyle+}    -   \sqrt{n(n-1)}\sin \vartheta_{n\scriptscriptstyle-}  \cos \vartheta_{n\scriptscriptstyle+}\right)   + \sqrt{n}\,\Gamma \sin \vartheta_{n\scriptscriptstyle+} \sin \vartheta_{n\scriptscriptstyle-}       =0,  
\end{equation}

The Landau levels of BLG with Rashba SOI are thus determinated by  the eigenvalues of  ${\tilde {\cal H}}_n$, which yields  
\begin{equation}
\varepsilon _{n,\mu \nu} =\mu \sqrt{ U^2+\left|Q_{n \nu}\right|^2 },
\end{equation}
\noindent  for $n\ge2$, being $n$ the Landau level index with  $\nu=\pm$ (plus/minus) state of the $\mu=\pm$  electron (hole) band. Using Eqs. (\ref{Qnbis}) along with (\ref{Nn})-(\ref{D2}), the formula (\ref{LL-BLG}) readily follows. The eigenvectors of $\tilde {\cal H}_n$ are given by
 
\begin{equation} 
\fl
\footnotesize
|\tilde\phi _{n1}\rangle=  \left(
\begin{array}{c}
 0  \\
 \cos\eta_{-} \\
 \sin\eta_{-} \\
 0
\end{array}
\right), \quad 
|\tilde\phi _{n2}\rangle=  \left(
\begin{array}{c}
 \cos\phi_{-} \\
 \sin\phi_{-} \\
 0 \\
0
\end{array}
\right), \quad
|\tilde\phi _{n3}\rangle=  \left(
\begin{array}{c}
 0  \\
  -\cos\eta_{+} \\
 \sin\eta_{+} \\
 0
\end{array}
\right), \quad 
|\tilde\phi _{n4 }\rangle=  \left(
\begin{array}{c}
 -\cos\phi_{+} \\
 \sin\phi_{+} \\
 0 \\
0
\end{array}
\right) ,
\end{equation}
  
\noindent where the angles $\eta_{\pm}$ and $\phi_{\pm}$ satisfy, 
\begin{equation}
\fl
\tan \eta_{\pm} =\frac{1}{\left|\,P_{\pm} \,\right|}=\frac{\left|\,Q_{n+}\,\right|}{\left|\,{ U}\pm\sqrt{U^2+Q_{n+}^2}\,\right|} , \quad\quad   {\rm and} \quad \quad \tan \phi_{\pm} = \frac{1}{\left|\,M_{\pm} \,\right|}=\frac{\left|\,Q_{n-}\,\right|}{\left|\,U\pm\sqrt{ U^2+Q_{n-}^2}\,\right|}, 
\end{equation}

\noindent with $M_-M_+ = P_+P_- =-1$.
The eigenvectors of ${\cal H}_n$ are thus given by $|   \phi_{nj} \rangle=\mathbb{  V}  |\tilde\phi_{nj}\rangle$, and consequently, the eigenvectors of $\hat H_n$ are finally determined by $|\psi_n\rangle = \left( \phi^{(1)}_{nj} \xi_{n-2},\,
\phi^{(2)}_{nj} \xi_{n-1},\,
\phi^{(3)}_{nj} \xi_{n },\,
\phi^{(4)}_{nj} \xi_{n+1} 
\right)^T$. The normalized eigenvectors for the states ($\pm$)  of the $n-$th Landau level of a given electron (hole) band $\mu$ are thus explicitly specified by
\begin{equation}
\fl
\label{LL1} 
|\psi _{n\mu }^{\scriptscriptstyle(+)}\rangle= \left(
\begin{array}{c}
 -i \sin \vartheta_{n\scriptscriptstyle +} \sin\eta_{\mu}\,\xi_{n-2}\\
   \cos \vartheta_{n\scriptscriptstyle +} \sin\eta_{\mu} \,\xi_{n-1}\\
 i  \sin \vartheta_{n\scriptscriptstyle -} \cos\eta_{\mu}\,\xi_{n }\\
 -  \cos \vartheta_{n\scriptscriptstyle -} \cos\eta_{\mu}\,\xi_{n+1}
\end{array}
\right) , \quad\quad 
|\psi _{n\mu }^{\scriptscriptstyle(-)}\rangle= \left(
\begin{array}{c}
  i \cos \vartheta_{n\scriptscriptstyle +} \sin\eta_{\mu}\,\xi_{n-2}\\
   \sin \vartheta_{n\scriptscriptstyle +} \sin\eta_{\mu}\,\xi_{n-1} \\
  -i  \cos \vartheta_{n\scriptscriptstyle -} \cos\varphi_{\mu}\,\xi_{n }\\
  -\sin\vartheta_{n\scriptscriptstyle -} \cos\varphi_{\mu}\,\xi_{n+1}
\end{array}
\right) ,
\end{equation}
From  Eqs. (\ref{LL1}) and the orthogonality of the oscillator wave functions $\xi_n$, it follows that  the components in the plane of BLG of both, the valley and   spin polarization vanishes identically for all Landau levels, $\langle {  \tau_x^{(n)} }  \rangle_{\mu \nu}=\langle {  \tau_y^{(n)} }  \rangle_{\mu \nu}=0$, and    $\langle  {  S^{(n)}_x}  \rangle_{\mu \nu}=\langle  {  S^{(n)}_x}  \rangle_{\mu \nu}=0$.  The valley polarization in the perpendicular direction (along the $z-$axis) reads 
\begin{equation}
\label{TauLL}
\langle { \tau_z^{(\nu)}} \rangle_{n,\mu} 
= \left\{ \begin{array}{cc} 
  -\cos(2\eta_{\mu}) \, ,&  {\rm for} \,\, \nu=+ \\
\sin^2\eta_{\mu}-\cos^2\varphi_{\mu} \,\,,\,\,   &  {\rm for}  \,\,\nu=-\,
\end{array}\right.
\end{equation}

 Note that it is $k-$independent and that in the limit case of zero bias voltage ($U=0$) results in $\eta_{\pm}=\phi_{\pm}=\frac{\pi}{4}$, and hence a zero valley polarization. 
The $z$-th component of the spin polarization gives on the other hand,
  \begin{equation}
\label{SLL}
\langle { S_z^{(\nu)}} \rangle_{n,\mu} 
= \left\{ \begin{array}{cc } 
-\cos  \vartheta_{n\scriptscriptstyle -} \cos^2\eta_{\mu}-\cos \vartheta_{n\scriptscriptstyle +} \sin^2\eta_{\mu} , & \nu=+ \\
 \cos  \vartheta_{n\scriptscriptstyle -} \cos^2\varphi_{\mu}+\cos  \vartheta_{n\scriptscriptstyle +} \sin^2\eta_{\mu},   & \nu=-  
\end{array}\right.
\end{equation}

\noindent which in the absence of bias voltage simplifies to Eq.(\ref{SznU0}).


\begin{thebibliography}{10}
\bibitem{comment1} 
The spectrum in single layer graphene with Rashba coupling is described by two zero-gap bands and two gapped bands splitted by 
$2\lambda_R$, Ref.\cite{Rashba}.
\end{thebibliography}
\end{document}